\documentclass{emulateapj}
\usepackage{lscape}
\shorttitle{2M0355: A Young, Nearby, Isolated Brown Dwarf.}
\shortauthors{Faherty et al.}

\begin{document}


\title{2MASSJ035523.37+113343.7: A Young, Dusty, Nearby, Isolated Brown Dwarf Resembling A Giant Exoplanet}
2MASSJ035523.37+113343.7

\author{Jacqueline K.\ Faherty\altaffilmark{1,2}, Emily L. Rice\altaffilmark{2,3}, Kelle L. Cruz\altaffilmark{2,4}, Eric E. Mamajek\altaffilmark{5,6}, Alejandro N\'{u}\~{n}ez\altaffilmark{2,4}}

\altaffiltext{1}{Department of Astronomy, 
Universidad de Chile Cerro Calan, Las Condes jfaherty17@gmail.com }
\altaffiltext{2}{Department of Astrophysics, 
American Museum of Natural History, Central Park West at 79th Street, New York, NY 10034; jfaherty@amnh.org }
\altaffiltext{3}{Department of Engineering Science \& Physics, College of Staten Island, 2800 Victory Blvd., Staten Island, NY 10301 USA}
\altaffiltext{4}{Department of Physics \& Astronomy, Hunter College, 695 Park Avenue, New York, NY 10065, USA}
\altaffiltext{5}{Cerro Tololo Inter-American Observatory, Casilla 603, La Serena, Chile}
\altaffiltext{6}{Department of Physics \& Astronomy, University of Rochester, Rochester, NY, 14627-0171, USA}
\begin{abstract}
We present parallax and proper motion measurements, near-infrared spectra, and WISE photometry for the low surface gravity L5$\gamma$ dwarf 2MASSJ035523.37+113343.7 (2M0355). We use these data to evaluate photometric, spectral, and kinematic signatures of youth as 2M0355 is the reddest isolated L dwarf yet classified. We confirm its low-gravity spectral morphology and find a strong resemblance to the sharp triangular shaped $H$-band spectrum of the $\sim$10~Myr planetary-mass object 2M1207b. We find that 2M0355 is underluminous compared to a normal field L5 dwarf in the optical and MKO $J,H$, and $K$ bands and transitions to being overluminous from 3-12 $\mu$m, indicating that enhanced photospheric dust shifts flux to longer wavelengths for young, low-gravity objects, creating a red spectral energy distribution.  Investigating the near-infrared color magnitude diagram for brown dwarfs confirms that 2M0355 is redder and underluminous compared to the known brown dwarf population, similar to the peculiarities of directly imaged exoplanets 2M1207b and HR8799bcd. We calculate UVW space velocities and find that the motion of 2M0355 is consistent with young disk objects ($<$ 2-3 Gyr) and it shows a high likelihood of membership in the AB Doradus association. 

\end{abstract}

\section{INTRODUCTION}
With masses intermediate between stars and planets (i.e., below the hydrogen burning and above the deuterium burning mass limit), brown dwarfs provide a natural link between stellar astrophysics and the planetary science of gas-giants (\citealt{Saumon1996}; \citealt{Chabrier1997}). Studies of the population have informed our understanding of low-mass star formation as well as the physical and chemical composition of low-temperature photospheres (e.g. \citealt{Burrows01,Burrows97}; \citealt{Chabrier00}). With an increasing number of brown dwarf discoveries, the diversity of the population in age, atmospheric properties, and chemical composition is becoming apparent.

Brown dwarfs are classified using red optical or near-infrared spectra and show characteristics which distinguish them as L (T$_{eff}\sim$2200 - 1300K) or T/Y (T$_{eff}<$1300) dwarfs (\citealt{Kirkpatrick99}; \citealt{Burgasser02}; \citealt{Cushing11}). The majority of spectrally classified field brown dwarfs within the literature are nearby isolated L dwarfs. Among the $\sim$1000 objects spanning this temperature regime, a significant portion exhibit near-infrared colors, spectral energy distributions (SEDs), and kinematics consistent with a field age population (e.g., \citealt{Kirkpatrick00}; \citealt{Knapp04}; \citealt{Cruz07}; \citealt{Chiu06}; \citealt{Faherty09}; \citealt{Schmidt10}). However there are subsets exhibiting strong deviations in observational properties from the general population including low-metallicity subdwarfs,  low surface gravity objects, and potentially cloudy/cloudless L dwarfs (\citealt{Burgasser03,Burgasser04,Burgasser07}; \citealt{Looper08}; \citealt{Cruz09}; \citealt{Cushing09}; \citealt{Kirkpatrick10}; \citealt{Rice10}; \citealt{Radigan12}). 

The most relevant sub-population to giant exoplanet studies are young (i.e.,  low surface gravity) isolated L dwarfs. The archetypal low surface gravity L dwarf, 2MASSJ01415823$-$4633574 (2M0141), was discovered by \citet{Kirkpatrick06}. Its optical spectrum exhibits strong bands of VO but abnormally weak TiO, K, and Na absorption. In the near-infrared, its red $J-K_{s}$ color (2MASS $J-K_{s}$=1.73) and triangular $H$-band spectral morphology distinguish it from field L dwarfs (\citealt{Kirkpatrick10}; \citealt{Patience12}). It is suspected to be a member of the $\beta$ Pictoris or Tucana-Horologium association, although the precise kinematics required to confirm association have not yet been determined (\citealt{Kirkpatrick10}). After the discovery and characterization of 2M0141, additional isolated L dwarfs sharing similar photometric and spectral peculiarities attributed to a  low surface gravity were reported (e.g. \citealt{Reid08}; \citealt{Cruz09}; \citealt{Kirkpatrick10}). While the ages of these seemingly young L dwarfs remain largely unconstrained, there are kinematic and spatial indications that they represent the lowest mass members of nearby moving groups such as AB Doradus, $\beta$ Pictoris, Tucana-Horalogium (\citealt{Cruz09}; \citealt{Kirkpatrick10}). 

\citet{Cruz09} point out that the majority of objects defining the population of the lowest surface gravity L dwarfs show spectral deviations indicating that they are younger than the Pleiades. Therefore using an age range\footnote{10 Myr chosen as the low-end range based on the age of the youngest nearby moving group.  100 Myr chosen as the upper limit based on an extrapolation and comparison to Pleiades age objects.} of $<$ 10-100 Myr and spectral classifications of early-mid type L dwarfs, these objects have masses close to--or in some cases below-- the deuterium burning limit, making them exoplanet analogs. Since young brown dwarfs are nearby and isolated, they are ideal laboratories for detailed studies of cool, low-gravity, dusty atmospheres that are similar to directly imaged exoplanets. 

In this paper we examine the kinematic, photometric, and spectral features of the  low surface gravity L5$\gamma$ dwarf 2MASSJ035523.37+113343.7 (2M0355). In section 2 we review published observations of 2M0355. In section 3 we describe new near-infrared spectral and imaging data, and in section 4 we evaluate indications of youth, including potential membership in nearby young moving groups. In section 5 we discuss the spectral energy distribution (SED) for 2M0355 as well as the near-infrared color-magnitude diagram for the brown dwarf population, highlighting the location of 2M0355 compared to directly imaged exoplanets. Conclusions are presented in section 6.

\section{Published Observations of 2M0355}
2M0355 was discovered by \citet{Reid06} in a search of the 2MASS database for ultracool dwarfs, but its observational peculiarities were not discussed until \citet{Reid08} and \citet{Cruz09}.  2M0355 is classified as an L5$\gamma$ dwarf\footnote{As suggested by \citet{Kirkpatrick05,Kirkpatrick06} and \citet{Cruz09} very low-gravity spectra are designated with subtype $\gamma$, intermediate gravity with $\beta$, and normal field objects with $\alpha$ (although $\alpha$ is typically omitted/implied for field objects.}, demonstrating strong Li absorption (EW 7.0$\AA$) and other signatures of low surface gravity in the optical (\citealt{Reid08}, \citealt{Cruz09}). Notably this source is the reddest isolated L dwarf yet classified, with a 2MASS $J-K_{s}$ color of 2.52$\pm$0.03. 

\citet{Reid06} examined 2M0355 for a close companion with the Near-Infrared Camera and Mutli-Object Spectrometer NIC1 on the \emph{Hubble Space Telescope} and found it unresolved. \citet{Blake07} examined this source for radial velocity variations but found no appreciable change over time and excluded the possibility of a companion with $M$ $sin$ $i$ $>$ 2.0~M$_{J}$ at any separation.  We note that \citet{Blake07} assumed an L dwarf primary mass of 100 M$_{J}$ which is large for even a field aged object, therefore, given the RV constraints, the limit is likely below 2.0 M$_{Jup}$. \citet{Bernat10} claimed the detection of a near-equal mass companion at 82.5 mas using aperture masking interferometry; however, this result falls at the low end of their confidence limits (90\%) and such a companion should have been detected by the \citet{Reid06} imaging campaign (although \citealt{Bernat10} note this object may be at the limit of \citealt{Reid06} detections).

Radial velocities of 12.24$\pm$0.18 and 11.92$\pm$0.22~km~s$^{-1}$ were measured by \citet{Blake07,Blake10}, respectively, using high-resolution $K$-band spectra from NIRSPEC on the Keck~II telescope and forward modeling techniques for high precision. Proper motion measurements have been reported in \citet{Schmidt07}, \cite{Casewell08}, and \cite{Faherty09}. We present an updated proper motion as well as a parallax in Section ~\ref{kinematics}.

\section{New Observations of 2M0355}
We obtained near-infrared spectroscopy and imaging of 2M0355 and report new low and medium resolution spectroscopy of the source as well as a parallax and improved proper motion measurements. 

\subsection{Near-Infrared Spectroscopy}
We obtained low- and medium resolution near-infrared spectroscopy using the SpeX spectrograph (\citealt{Rayner03}) mounted on the 3m NASA Infrared Telescope Facility (IRTF). On 2007 November 13, we used the spectrograph in cross-dispersed mode (SXD) with the 0$\farcs$5 slit aligned to the parallactic angle to obtain $R~\equiv~\lambda$ / $\Delta\lambda~\approx$~1200 spectral data over the wavelength range of 0.7--2.5 $\mu$m. The conditions of this run were clear and stable with seeing of 0$\farcs$5 at $K$. We obtained 6 individual exposure times of 300 seconds in an ABBA dither pattern along the slit. 

On 2011 December 7, we used the spectrograph in prism mode with the 0$\farcs$5 slit aligned to the parallactic angle. This resulted in $R~\equiv~\lambda$ / $\Delta\lambda~\approx$~120 spectral data over the wavelength range of 0.7--2.5 $\mu$m. Conditions included light cirrus and the seeing was 0$\farcs$8 at $K$. We obtained 10 individual exposure times of 90 seconds in an ABBA dither pattern along the slit.  Table ~\ref{observing} contains details on all observations reported in this work.

 Immediately after the science observation we observed the A0V star HD~25175 (Prism mode) or HD 25258 (SXD mode) at a similar airmass for telluric corrections and flux calibration. Internal flat-field and Ar arc lamp exposures were acquired for pixel response and wavelength calibration, respectively. All data were reduced using the SpeXtool package version 3.4 using standard settings (\citealt{Cushing04}, \citealt{vacca03}).

\subsection{Near-Infrared Imaging}
We observed 2M0355 with the Infrared Side Port Imager (ISPI, \citealt{van-der-Bliek04}) on the CTIO 4m Blanco telescope six times between 2008 October 11 and 2012 February 05. All observations used the $J$ band filter, under seeing conditions up to 2$\arcsec$ full width half maximum (FWHM) with typical conditions between 0.8--1.1$\arcsec$. ISPI has an $\sim$ 8 arcminute field of view and plate scale of 0.303$\arcsec$ per pixel. At each epoch and depending on the conditions, 5-10 images with 10-30~s and 2-4 co-adds were obtained while the target was $\pm$30 minutes off the meridian (Table ~\ref{observing}). Dark frames and lights on/off dome flats were obtained at the start of each evening. We used the Carnegie Astrometric Planet Search software to extract all point sources from each epoch and solve for relative parallaxes and proper motions (\citealt{Boss09}). The full image reduction procedures as well as the description of the parallax pipeline are described in \citet{Faherty12}.

\section{Evaluating Youth Indicators\label{youth}}
Youth indicators for isolated L dwarfs are not yet fully quantified or calibrated, but a number of distinguishing characteristics have been extrapolated from low-mass members (primarily late-type M dwarfs) of nearby young moving groups, open clusters and star forming regions or companions to young stars and confirmed by low-gravity atmosphere models (e.g. \citealt{Lucas01};  \citealt{Gorlova03}; \citealt{Luhman04}; \citealt{McGovern04}; \citealt{Allers07}; \citealt{Rice10,Rice11}, \citealt{Patience12}). 

Among the strongest indicators is the shape of the near-infrared spectra of young brown dwarfs which are subtly different than those of their field counterparts. Known brown dwarf members of the Chamaelleon II, Ophiuchus, Orion Nebula Cluster, TW Hydrae, and $\beta$ Pictoris groups demonstrate various degrees of sharply peaked $H$-band spectra compared to field aged objects.  The shape of the near-infrared continuum induced by steam absorption is sensitive to an objects surface gravity; therefore at younger ages, hence lower gravities, the $H-$band spectrum is peaked (\citealt{Luhman04} and Figure 6 from \citealt{Rice11}). 

An equally important indicator for brown dwarf members of young groups is a strong deviation in near-infrared color (significantly redder J-K$_{s}$) from the mean of a given spectral subtype. The clearest example is 2MASS J12073346$-$3932539 (2M1207b), a late-L dwarf member of the TW Hydrae association with $J-K$=3.05, $\sim$0.5 mag redder than any other known L dwarf (\citealt{Chauvin04}; \citealt{Mohanty07}). Similar to the spectral deviations of young brown dwarfs, the photometric peculiarities can be explained as a consequence of lower surface gravity. At lower values--hence lower pressure at a given temperature in the photosphere--, H$_{2}$ collision induced absorption (CIA) is reduced leading to a reduction of the strongest absorption feature at 2.5 $\mu$m (less absorption at $K$ band relative to $J$ band) and a red $J-K$ color (\citealt{Kirkpatrick06}).   An evolutionary model comparison of a large collection of low-surface gravity or young companion brown dwarfs to tracks with differing cloud, metallicity and gravity properties demonstrates that the change in near-IR color is attributed to changes in CIA H$_{2}$ affected by lower-surface gravities (see \citealt{Faherty12} and references there-in).

Additionally, the kinematics of young brown dwarfs as a population can be used as an indicator of youth as they are distinctly different from the kinematics of the field brown dwarf population. As discussed in \citet{Faherty09,Faherty12} low surface gravity brown dwarfs have significantly smaller tangential velocities and dispersions than the overall brown dwarf population. The young age (likely $<$ 1 Gyr) of these sources means they have spent less time in the disk so they have had minimal interactions with nearby stars and giant molecular clouds that will eventually increase their overall velocity dispersion (e.g. \citealt{Weinberg87}; \citealt{Faherty10}; \citealt{Dhital10}). 
In the following subsections we compare the photometry, near-infrared spectral features, and kinematics of 2M0355 to known young brown dwarfs, directly-imaged exoplanets, and the field population in order to evaluate signatures of youth for this unusual object.

\subsection{Photometry\label{Photometry}}
\begin{figure*}[!ht]
\begin{center}
\epsscale{1.0}
\plotone{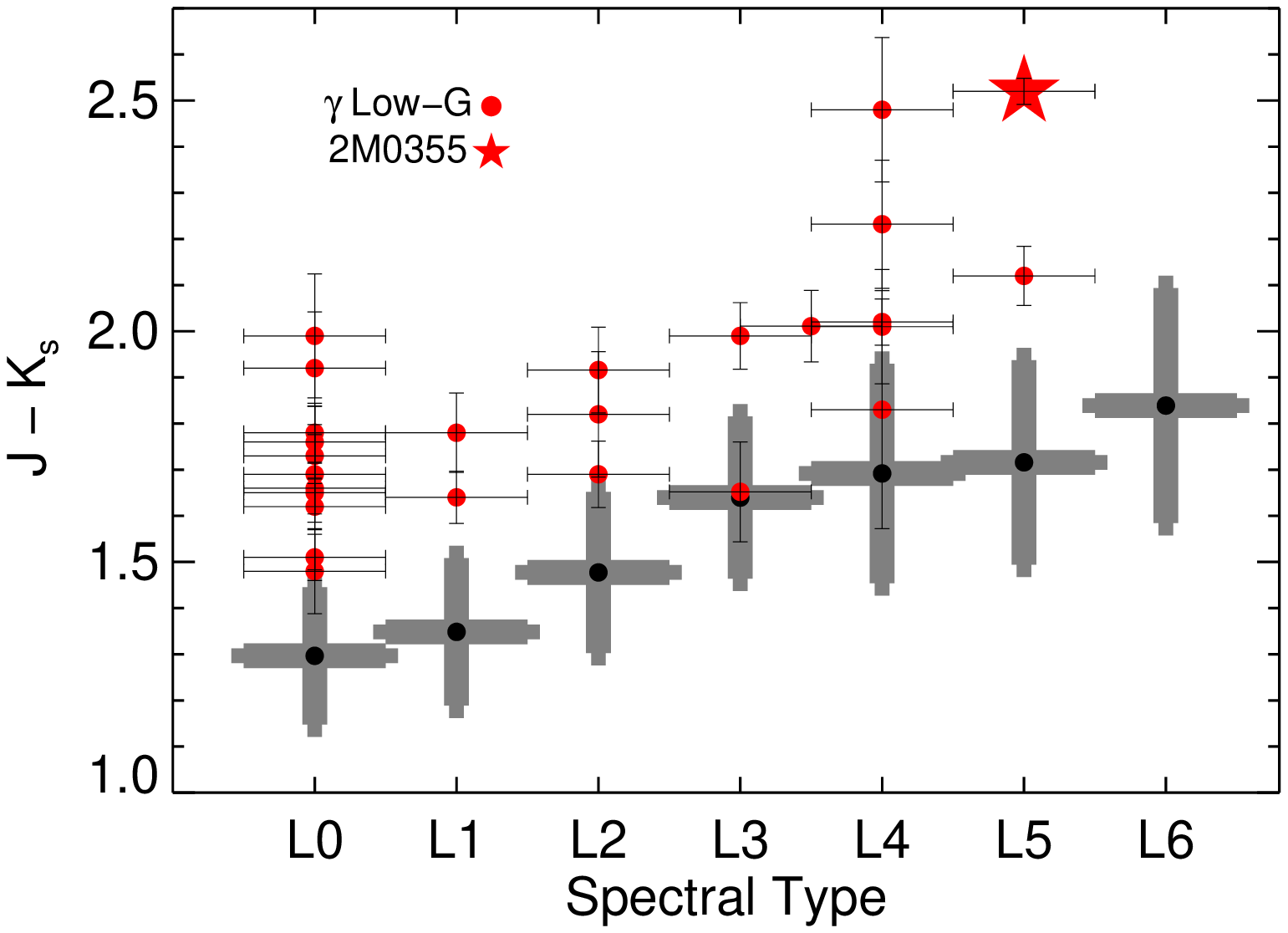}
\end{center}
\caption{2MASS (J-K$_{s}$) color versus spectral type for field L dwarfs. Mean colors of normal (excluding subdwarfs, young, and  low surface gravity) objects are displayed as grey bars and listed in Table~\ref{meancolors}.  Only sources with $J$ or $K$ uncertainties $<$ 0.1 are used. Low surface gravity $L\gamma$ dwarfs are red filled circles and are listed in Tables~\ref{low-G}-~\ref{colorslow-G}. 2M0355 is marked as a five-point star.  } 
\label{fig:jmk}
\end{figure*}

2M0355 is the reddest isolated L dwarf known. In Figure ~\ref{fig:jmk} we show the mean $J-K_{s}$ color and standard deviation for L dwarfs (binned by 0.5 subtype) calculated from a compilation of field objects\footnote{The compiled list of L dwarfs comes primarily from the DwarfArchives.org combined with the results of \citet{Schmidt10}.} with photometric uncertainties $<$ 0.1, excluding known young objects and subdwarfs. For comparison, other confirmed low-gravity L$\gamma$ dwarfs are plotted as filled circles and 2M0355 as a filled five-point star. In Table~\ref{meancolors} we list the average infrared photometric properties of field L dwarfs, and in Tables~\ref{low-G} and~\ref{colorslow-G} we list the infrared photometry and colors of low-gravity L dwarfs, respectively.

With a $J-K_{s}$ color of 2.52$\pm$0.03, 2M0355 is 0.8 mag redder than the average for L5 dwarfs, or nearly 4$\sigma$ from the mean color. A similar deviation from the mean of the subtype is seen among other low surface-gravity L$\gamma$ dwarfs listed in Tables~\ref{low-G}-~\ref{colorslow-G}, but 2M0355 is the most extreme example (although we note that the L4 dwarf 2MASSJ1615+4953 shows very similar  deviations in both its J-K$_{s}$ and W1-W2 colors). As discussed above,  low surface gravity effects leading to a reduction in H$_{2}$ collision induced absorption is the likely cause for the extreme deviation. However we note that not all unusually red L dwarfs demonstrate low surface gravity spectral features; therefore this peculiarity alone is not conclusive about age (e.g. \citealt{Looper08}).

\begin{figure*}[!ht]
\begin{center}
\epsscale{1.0}
\plotone{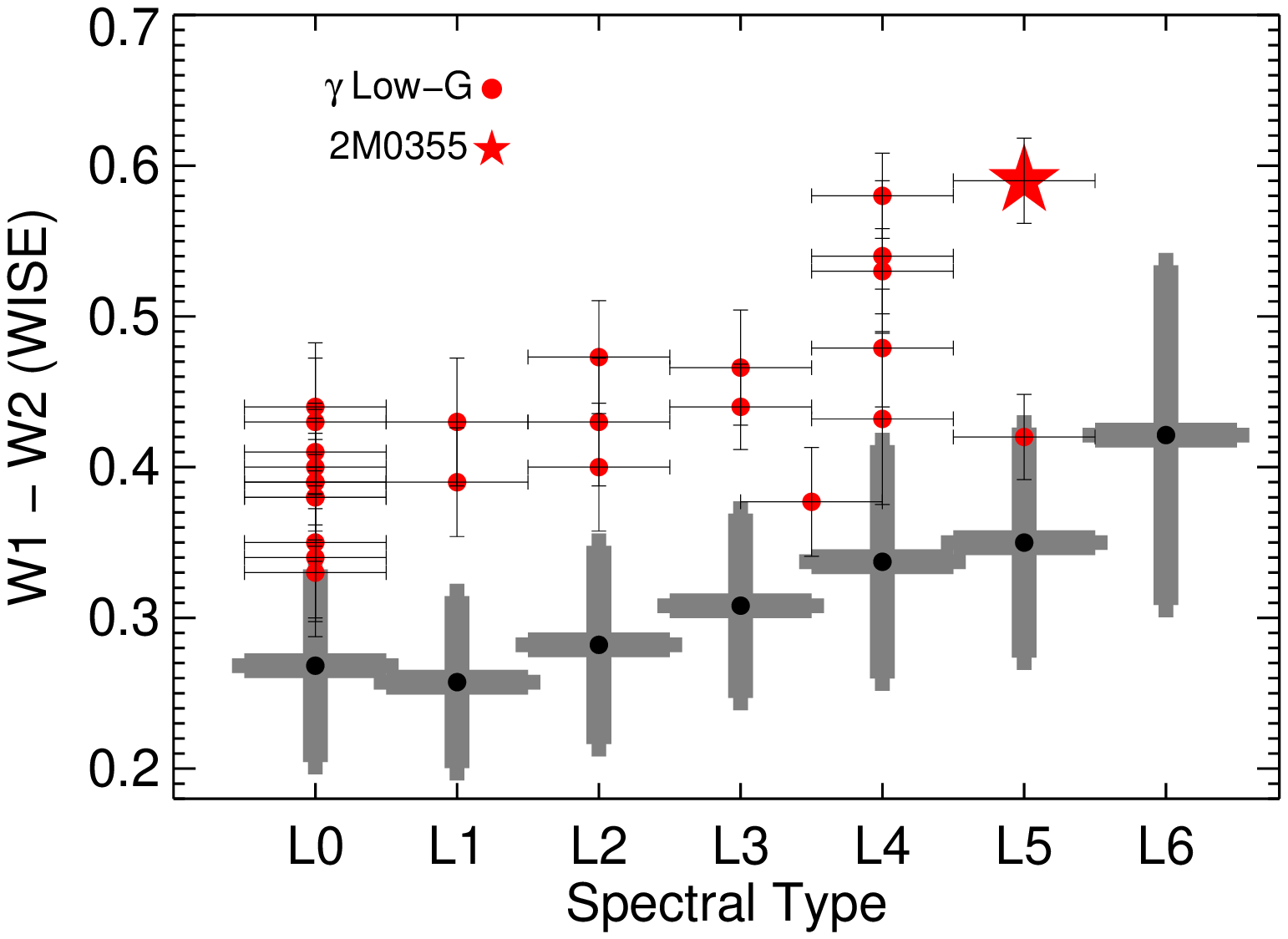}
\end{center}
\caption{WISE (W1-W2) color versus spectral type for field L dwarfs. Mean colors of normal (excluding subdwarfs, young, and  low surface gravity) objects are displayed as grey bars and listed in Table~\ref{meancolors}.  Only sources with $W1$ or $W2$ uncertainties $<$ 0.1 are used. Low surface gravity $L\gamma$ dwarfs are red filled circles  and are listed in Tables~\ref{low-G}-~\ref{colorslow-G}. 2M0355 is marked as a five-point star.} 
\label{fig:w1mw2}
\end{figure*}

In the same manner as Figure~\ref{fig:jmk} we compile WISE photometry of known field L dwarfs with photometric uncertainties $<$ 0.1, excluding subdwarfs and confirmed young objects, to calculate the mean $W1-W2$ color and corresponding standard deviation for spectral subtypes (again binned by 0.5 subtype) and highlight the photometry of 2M0355 (see also Table~\ref{meancolors}). As demonstrated in Figure~\ref{fig:w1mw2}, with a $W1-W2$ color of 0.59, 2M0355 is 0.24 mag redder than the average of its spectral subtype or 3$\sigma$ from the mean color. Comparing with the 25 similarly classified L$\gamma$ dwarfs, we find that 2M0355 is the reddest known isolated L dwarf in near and mid-infrared colors.

\subsection{Spectral Features\label{spectra}}
\begin{figure*}[!ht]
\begin{center}
\epsscale{1.0}
\plotone{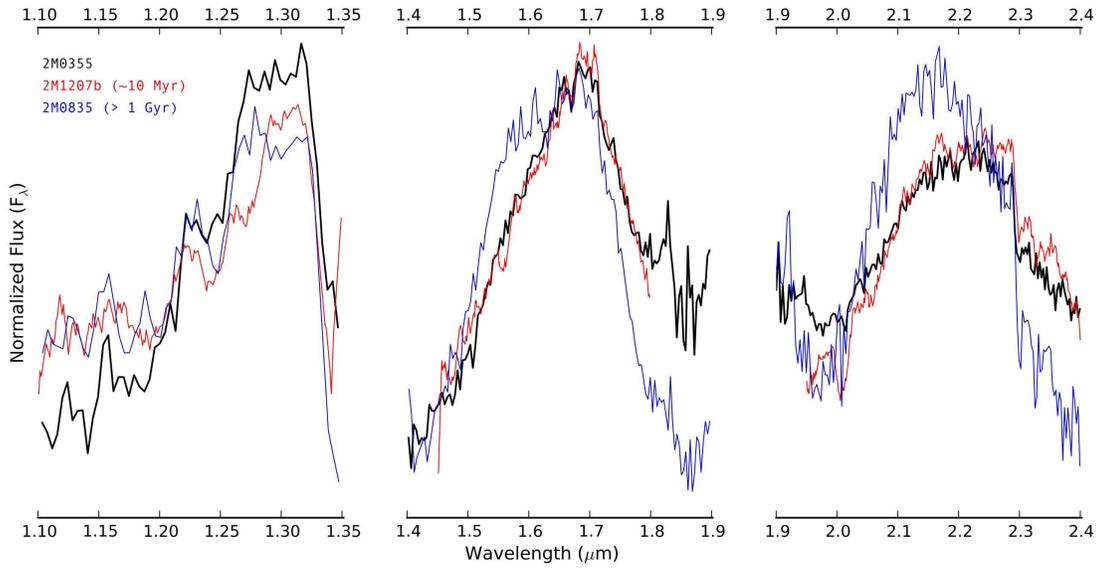}
\end{center}
\caption{The SpeX prism near-infrared spectra of 2M0355 (black solid line) compared to the L5 near-infrared standard (blue dashed line) 2M0835 (defined in \citealt{Kirkpatrick10}) and the young planetary mass companion (red dashed line) 2M1207b (from \citealt{Patience10}). We separate $JHK$ bands and normalize the three objects over each band independently.  2M0355 deviates from the field L5 with a sharply peaked $H$ band and suppressed $K$ band, and matches well with the features of 2M1207b.} 
\label{fig:SpeX2}
\end{figure*}

2M0355 is classified as an L5$\gamma$ in the optical by \citet{Cruz09} based on its similarity to field L5's but with very weak FeH absorption and weak Na~{\sc I} and K~{\sc I} lines, which are typically interpreted as signatures of low surface gravity. In Figure~\ref{fig:SpeX2} we show the SpeX prism spectrum for 2M0355 and compare it to the field L5 (presumed age $>$ 1 Gyr) near-infrared standard 2MASSJ08350622+1953050 (2M0835) as well as the $\sim$10 Myr L dwarf 2M1207b (\citealt{Chiu06}; \citealt{Kirkpatrick10}; \citealt{Patience12}). We normalize the spectra separately in each bandpass and smooth 2M1207b by a factor of 3. The shape of 2M0355 in all three bands deviates significantly from the spectrum of the field standard.  Compared to 2M1207b, the $H$ and $K$ bands are very similar, but the $J$ band is intriguingly different. 2M0355 has a steeper slope from 1.1-1.25 $\mu$m and a wider peak at~1.30$\mu$m that is more similar to the field object.  In a forthcoming paper, we will present a detailed $J$ band spectral analysis of  2M0355 and other young brown dwarfs compared to their field counterparts.

\begin{figure*}[!ht]
\begin{center}
\epsscale{1.0}
\plotone{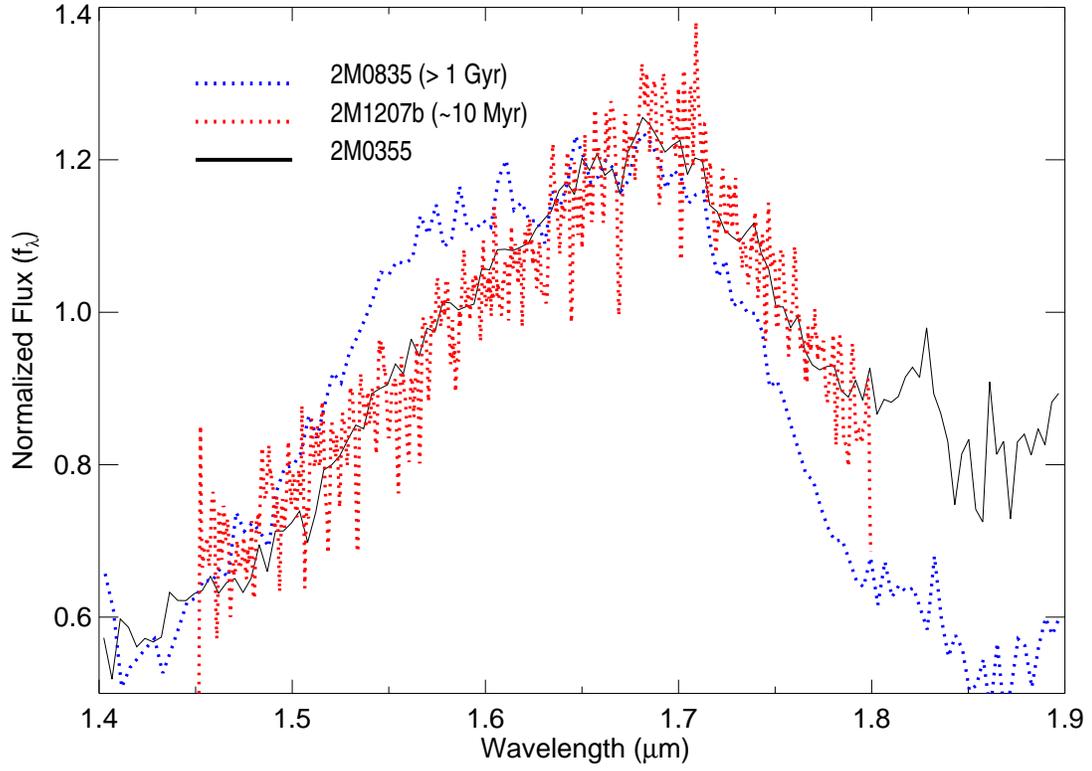}
\end{center}
\caption{SpeX cross-dispersed $H$-band spectra of 2M0355 (black solid line) compared to the field L5 near-infrared standard (blue dashed line) 2M0835  (defined in \citealt{Kirkpatrick10}) and the young planetary mass companion (red dashed line) 2M1207b (from \citealt{Patience12}). The strong triangular shape seen in 2M1207b and 2M0355 is interpreted as a hallmark of low surface gravity.} 
\label{fig:SpeX}
\end{figure*}

Several near-infrared spectral features are sensitive to surface gravity, including the $H$-band where a sharp triangular peak is seen consistently for known young brown dwarfs at a range of ages (e.g. \citealt{Lucas01};   \citealt{Luhman04}; \citealt{Allers07}; \citealt{Rice10,Rice11}). In Figure~\ref{fig:SpeX}, we present higher-resolution ($R\sim$1200) $H$-band spectra of 2M0355 as well as the same comparative objects shown in Figure~\ref{fig:SpeX2}. There is an excellent match between the sharp peak of 2M1207b and 2M0355, distinct from the plateau at $\sim$1.55--1.70~$\mu$m of the field object. Combined with the photometric peculiarities, this is a strong indicator that 2M0355 is significantly younger than the field object ($<<$ 1 Gyr).

\subsection{Kinematics\label{kinematics}}
\begin{figure*}[!ht]
\begin{center}
\epsscale{1.0}
\plotone{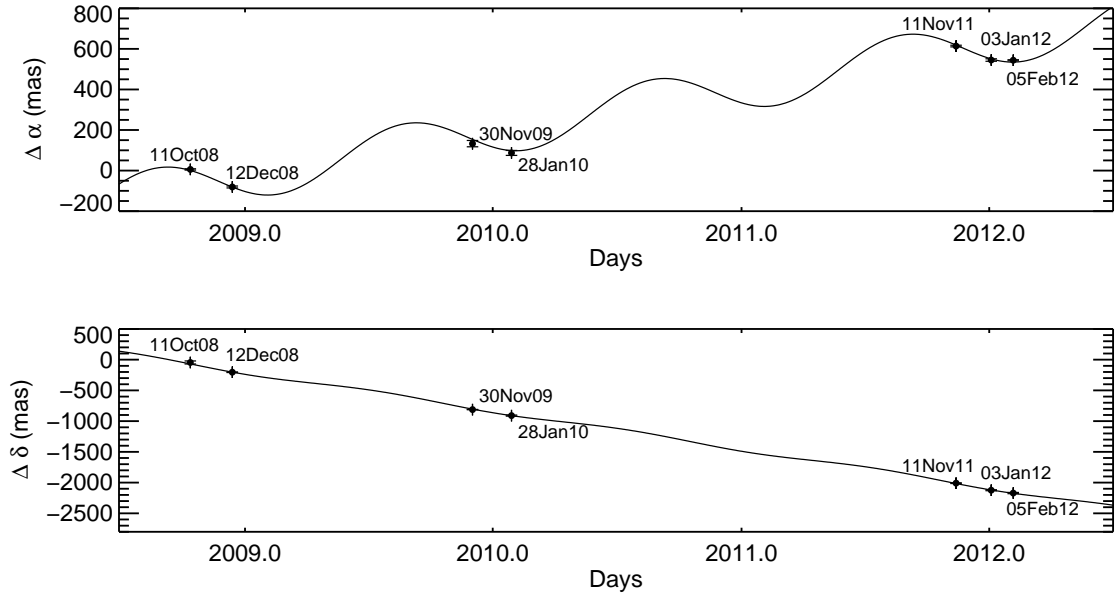}
\end{center}
\caption{ISPI astrometry for 2M0355.  Upper panel shows the imaging results (filled points) as well as the best fit solution for the proper and parallactic motion in right ascension and the lower panel shows the same in declination.  } 
\label{fig:astrometry}
\end{figure*}

Using multi-epoch ISPI data (see Figure ~\ref{fig:astrometry}), we report improved proper motion and parallax measurements for 2M0355. The proper motion was measured previously by \citet{Schmidt07}, \cite{Casewell08}, and \cite{Faherty09}. Our updated value is consistent with previous values but with 50-60$\%$ smaller error bars. The new parallax measurement of $\pi_{abs}$=122$\pm$13~mas\footnote{We measure $\pi_{rel}$=120$\pm$ 12 ~mas with a 2~mas correction from relative to absolute astrometry.}  for 2M0355 places the L5$\gamma$ dwarf at a distance of 8.2$^{+1.0}_{-0.8}$~pc. We list all astrometric and photometric properties in Table ~\ref{properties}.

\subsection{Moving Group Membership}
\begin{figure*}[!ht]
\begin{center}
\epsscale{1.0}
\plotone{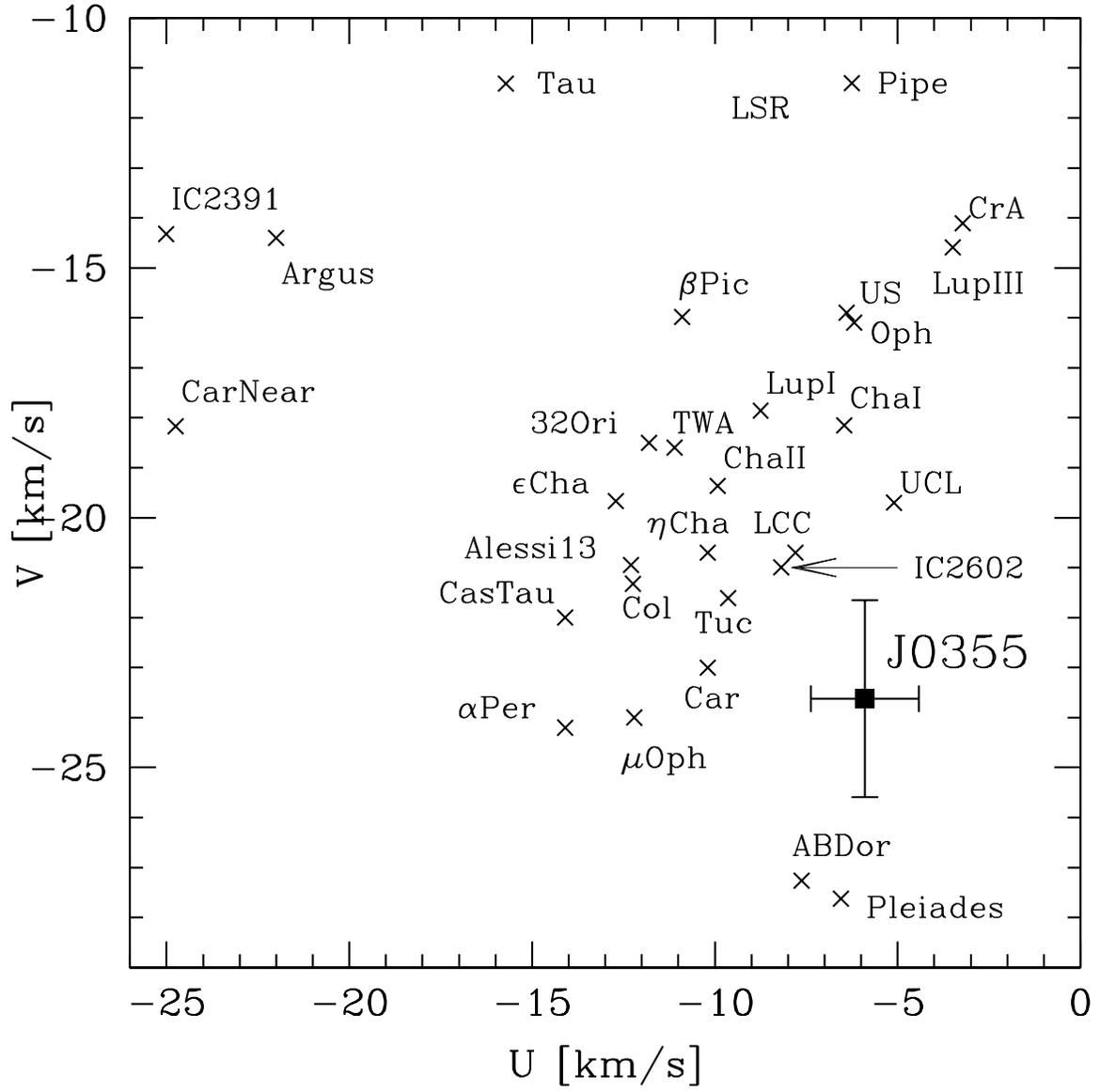}
\end{center}
\caption{The U versus V velocity plot for a large collection of young stellar groups within 200 pc of the Sun from \citet{Mamajek10}.  2M0355 is highlighted as a filled square and is located in a busy region of velocity space for nearby young objects.   } 
\label{fig:space}
\end{figure*}

\begin{figure*}[!ht]
\begin{center}
\epsscale{1.2}
\includegraphics[width=.55\hsize]{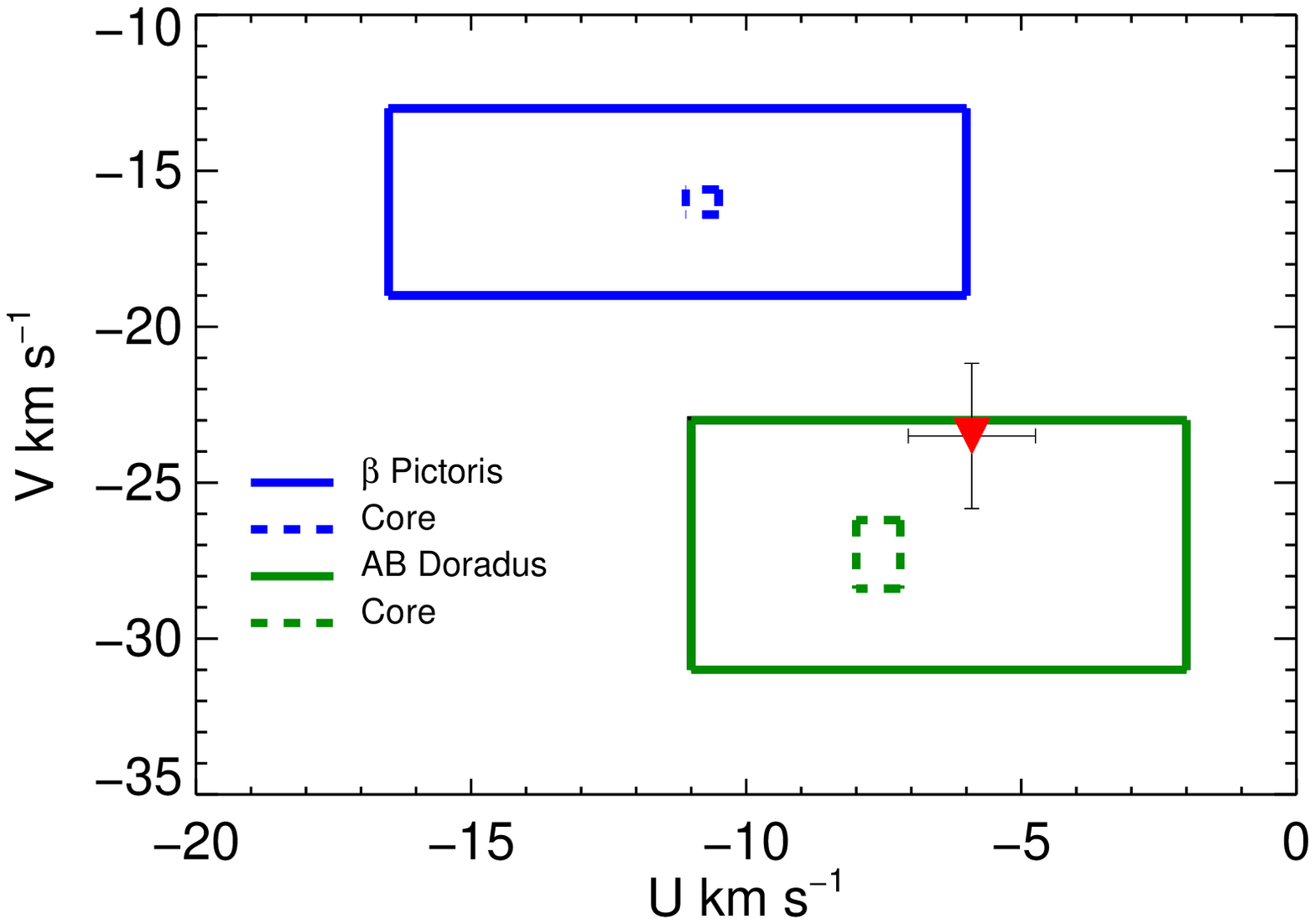}
\plottwo{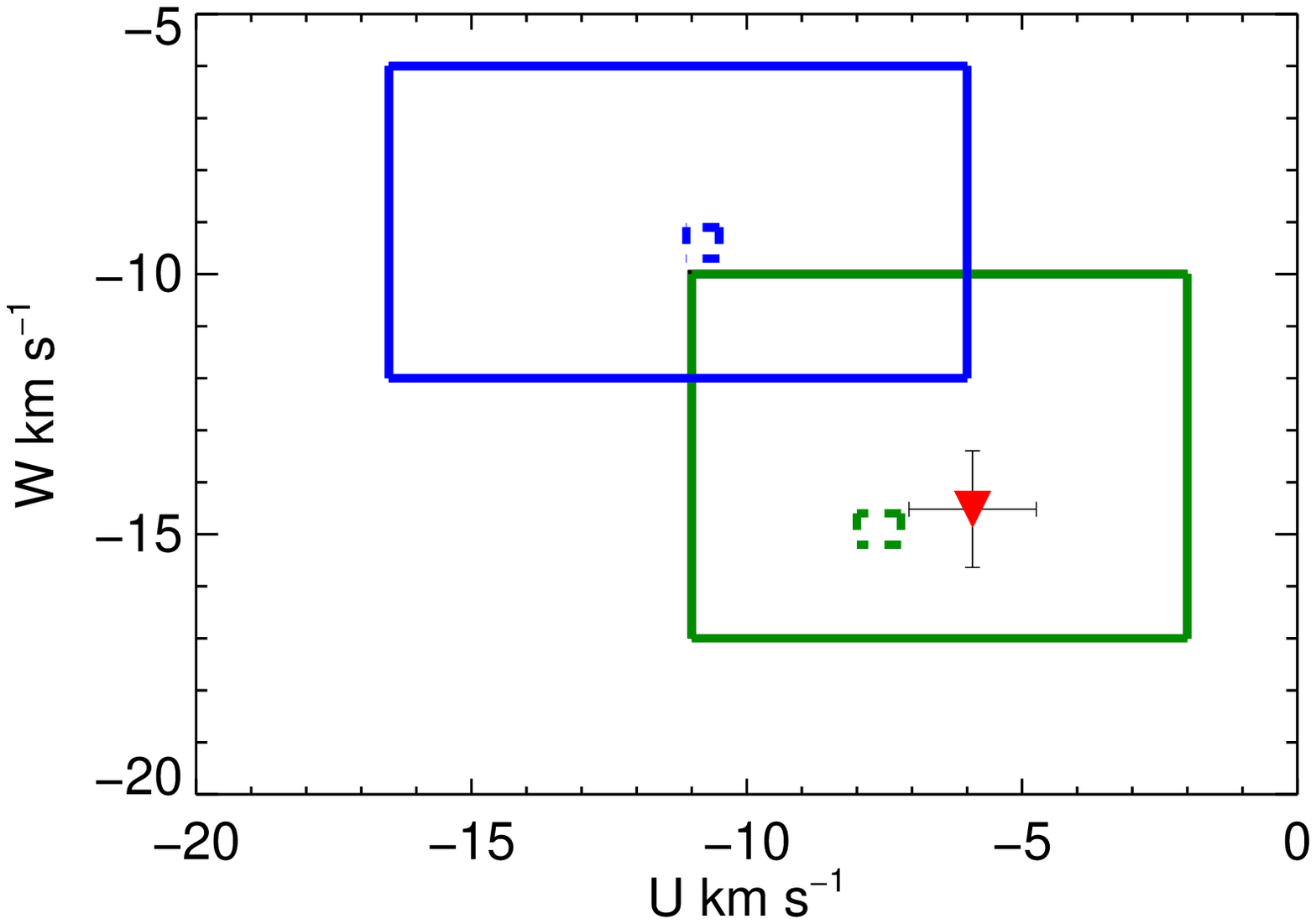}{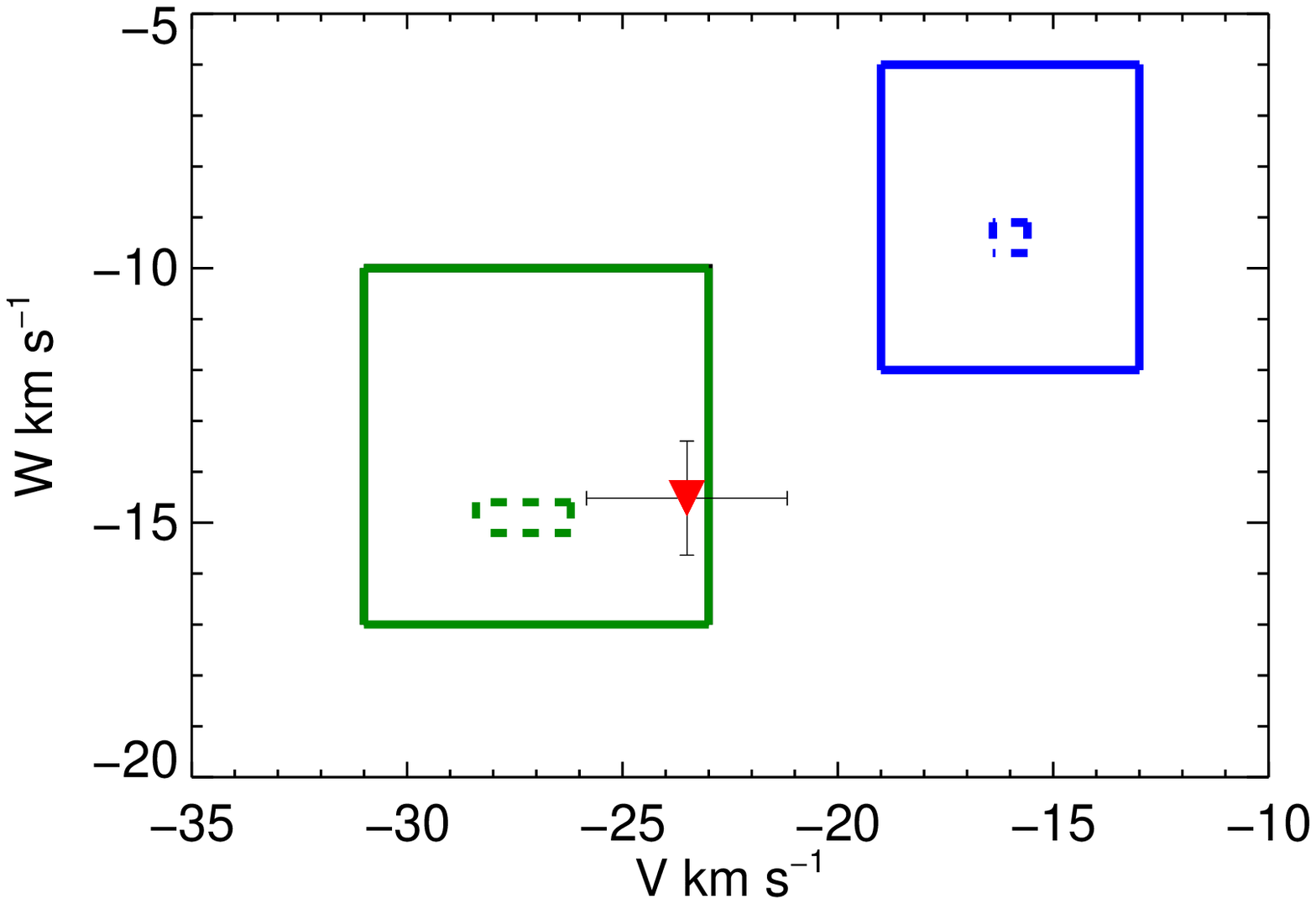}
\end{center}
\caption{The UVW properties of 2M0355 (red filled triangle) compared to those of the members of the nearby young groups $\beta$ Pictoris (blue) and AB Doradus (green). Solid rectangles surround the furthest extent of highly probable members from \citet{Torres08} but their distribution does not necessarily fill the entire rectangle. We also show the core UVW values of $\beta$ Pictoris ($-$10.9$\pm$0.3, $-$16.0$\pm$0.3, $-$9.2$\pm$0.3) km s$^{-1}$ and AB Doradus ($-$7.6$\pm$0.4, $-$27.3$\pm$1.1, $-$14.9$\pm$0.3) km s$^{-1}$  from \citet{Mamajek10} updated using the revised Hipparcos astrometry from \citet{van-Leeuwen07} and compiled velocity catalog of \citet{Gontcharov06}.  
} 
\label{fig:kinematics}
\end{figure*}

At a distance of 8.2~pc and with spectral and photometric differences from the field population resembling those of the $\sim$10 Myr 2M1207b, we investigate whether 2M0355 could be kinematically associated with one of the nearby young moving groups. Using the proper motion and parallax measured in this work with the most recent radial velocity from \citet{Blake10},  we calculate ($U,V,W$) = ($-$5.9$\pm$1.5, $-$23.6$\pm$2.0, $-$14.6$\pm$1.3) km s$^{-1}$ for 2M0355\footnote{UVW values are calculated in a left-handed coordinate system with $U$ positive toward the Galactic center.}. These calculated space velocities are consistent with thin disk membership (age $<$ $\sim$2-3 Gyr), and the tangential velocity of 21.5$\pm$1.2 km s$^{-1}$ is consistent with the population of low gravity, kinematically young brown dwarfs (\citealt{Faherty09,Faherty12}; \citealt{Eggen89a,Eggen89}).  In Figure ~\ref{fig:space} we show the $UV$ velocities for a number of young stars or clusters within 200~pc of the Sun and find that 2M0355 is at the edge of a well populated region of velocity space. Figure~\ref{fig:kinematics} shows Galactic space velocities compared to $\beta$ Pictoris, and AB Doradus, the two closest moving groups to the Sun and the most likely groups of which 2M0355 might be a member. We find that 2M0355 overlaps within 1$\sigma$ of the range in UVW values for probable members of AB Doradus.

To examine the likelihood of 2M0355's membership in nearby moving groups, we determine the $\chi^{2}$ probability for several known stellar groups within 75pc. We include a field star model and nearby moving group parameters from \citet{Malo12} and supplement with the parameters for the Ursa Major, Hyades, and Carina Near groups.  For most groups, we adopt the centroid positions and dispersions calculated by \citet{Malo12}, however we use velocity estimates either calculated by us or from the recent literature, where we split the uncertainties in the centroid velocities from their 1D intrinsic velocity dispersions \footnote{We
adopt the following parameters throughout the analysis (centroid velocities and standard errors, followed by centroid positions and $1\sigma$ dispersions): Ursa Major: $(U, V, W)$ = (15.0, 2.8, -8.1) $\pm$ (0.4, 0.7, 1.0)
km s$^{-1}$ and $(X, Y, Z)$ = (-4.4, 6.2, 18.2) $\pm$ (16.7, 15.4, 17.0) pc (calculated using membership from \citet{Madsen02}).  Carina Near: $(U, V, W)$ = (-24.8,-18.2, -2.3) $\pm$ (0.7, 0.7, 0.4) km s$^{-1}$ and $(X, Y, Z)$ = (0.1, -31.7, -9.2) $\pm$ (4.3, 5.6, 1.1) pc (calculated using membership from \citet{Zuckerman06}). 
Hyades: $(U, V, W)$ = (-42.3, -19.1, -1.5) $\pm$ (0.1, 0.1, 0.2) km s$^{-1}$ and $(X, Y, Z)$ = (-43.0, 0.3, -17.3) $\pm$ (3.8, 3.5, 3.1) pc.  For the TWA group we adopt the recent centroid velocity from \citet{Weinberger11} of $(U, V, W)$ = (-11.1, -18.6, -5.1) $\pm$ (0.3, 0.2, 0.2) km s$^{-1}$.  Based on unpublished calculations by \citealt{Mamajek10}, in prep) we adopt intrinsic 1D velocity dispersions of 1.0 km s$^{-1}$ for AB Doradus, 1.1 km s$^{-1}$ for Tucana-Horologium, 1.3 km s$^{-1}$ for Carina Near, 1.5 km s$^{-1}$ for Ursa Majoris, and $\beta$ Pictoris, 0.8 km s$^{-1}$ for TWA (Mamajek 2005), and 1 km s$^{-1}$ for the Hyades and Argus.}.

We first determine a $\chi^{2}$ probability that estimates the percentage of real members of a given group expected to have $\chi^{2}$ values higher than that of 2M0355--allowing for 2M0355's observational errors and the estimated intrinsic velocity spread and spatial distribution of group members. Then we calculate a ``final" probability, normalizing by the sum of the individual (marginal) star-group probabilities. At this time, equal weights are assigned to the field star and individual group models (further refinement would be beyond the focus of this study).

The initial $\chi^{2}$ probability for 6 degrees of freedom is calculated as:

\begin{equation}
\chi^{2} =A+B
\end{equation}

\begin{equation}
A=\frac{(U_{o} - U_{g})^{2}}{\sigma^{2}_{U}} + \frac{(V_{o} - V_{g})^{2}}{\sigma^{2}_{V}} + \frac{(W_{o} - W_{g})^{2}}{\sigma^{2}_{W}}
\end{equation}

\begin{equation}
\sigma_{i}=\sqrt{\sigma_{i,o}^{2}+\sigma_{i,g}^{2}+\sigma_{i,d}^{2}}
\end{equation}

\noindent where $i$ is indexed as $U,V$ or $W$, $o$ is the component for 2M0355, $g$ is the component of the group, and $d$ is the intrinsic 1-D $i$-velocity dispersion of the group.

\begin{equation}
B=\frac{(X_{o} - X_{g})^2}{\Delta_{X}^{2}} + \frac{(Y_{o} - Y_{g})^{2}}{\Delta_{Y}^{2}} + \frac{(Z_{o} - Z_{g})^2}{\Delta_{Z}^{2}}
\end{equation}

\begin{equation}
\Delta_{j} = \sqrt{\Delta_{j,o}^{2} + \Delta_{j,g}^{2}}
\end{equation}

\noindent where $j$ is indexed as $X,Y$, or $Z$, $\Delta_{j}$ is defined as the 1$\sigma$ dispersion in the Galactic
cartesian coordinates; $o$ is the component for 2M0355, $g$ is the component for the group  (we ignore the uncertainties in the group
centroids which are negligible compared to the 1$\sigma$ dispersions).

Using this method, we estimate that 73\% of AB Doradus members
would have velocities and positions more discrepant than that for
2M0355, while only 0.06\% of $\beta$ Pictoris members would have more discrepant
values. Approximately 99.9\% of field stars would
have velocities and positions more discrepant than that of 2M0355, although
this is likely skewed by the fact that the field star centroid (as well as our source) is so close to
the Sun. 

The $\chi^{2}$ probabilities for the other groups
investigated within 75 pc (Ursa Majoris, Carina Near, Tucana Horologium, Hyades, Argus, TWA),
all yielded probabilities of $<$10$^{-17}$\%. If one sums the
individual marginal group and field star membership probabilities and
assigns equal weights, then we estimate that 2M0355 has a 42\% chance
of being an AB Doradus member, a 58\% chance of being a field star, and a
$<$0.04\% chance of being a $\beta$ Pictoris group member. Further work
calculating the relative densities of the young stellar groups could
refine these probabilities, but at this point it appears most
plausible that 2M0355 is either a member of the AB Doradus moving group or
a field star.  Given the
photometric and spectroscopic evidence for youth shown here-in combined with the low density of very young field stars, we believe
that the kinematic evidence points towards 2M0355 being a likely member
of the AB Doradus group.

  \section{Discussion}
Among the known population of low surface gravity L dwarfs, 2M0355+11 is the latest spectral type or one of the coolest isolated young brown dwarfs known.   To extend the comparison of young brown dwarfs and planetary-mass objects, we investigate the colors and luminosities of 2M1207b and the directly imaged planets HR~8799bcd. 

\begin{figure*}[!ht]
\begin{center}
\epsscale{1.5}
\plottwo{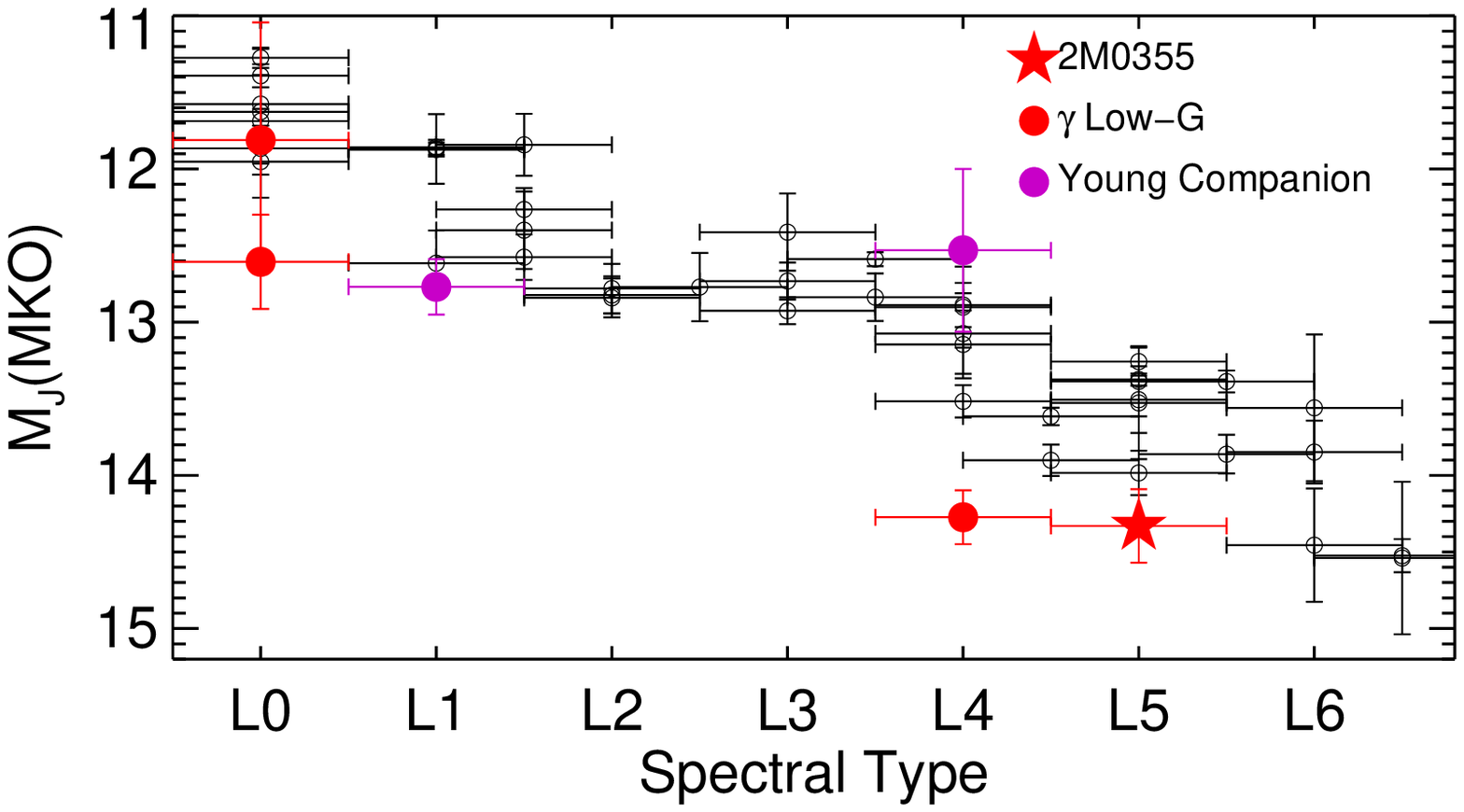}{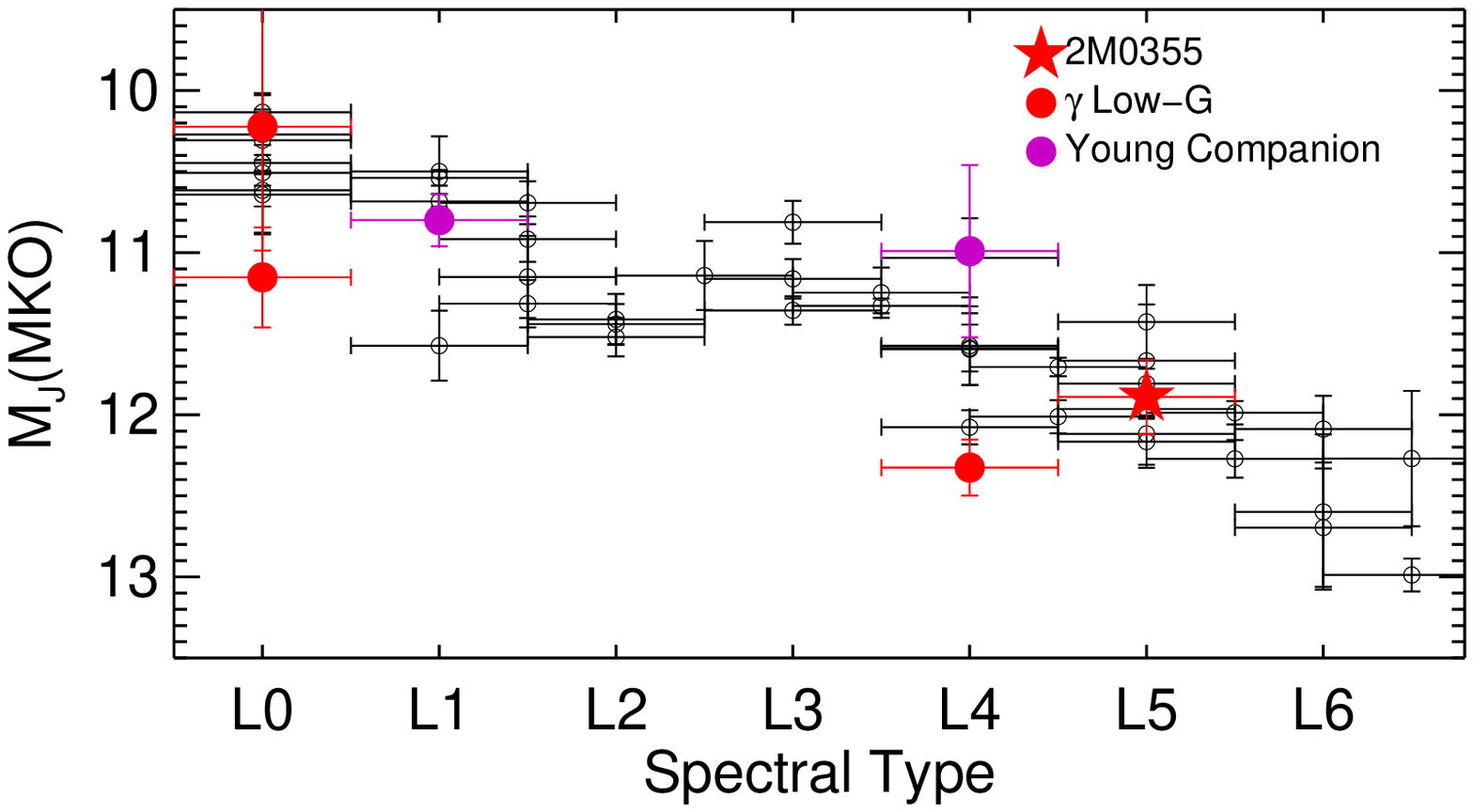}
\end{center}
\caption{The spectral type versus absolute magnitude diagram in MKO J (top) and K (bottom) for L dwarfs.  Normal objects (non binary, young or subdwarf) are shown as open black circles, L$\gamma$ dwarfs as filled red circles, and young companion brown dwarfs as filled purple circles.  2M0355 is highlighted as a red five-point star.} 
\label{fig:spt}
\end{figure*}

We calculated the absolute magnitude of 2M0355 from the new parallax as well as Mauna Kea Observatory  (MKO; \citealt{Tokunaga02}) apparent magnitudes converted from 2MASS photometry using the \citet{Stephens04} relations.  Comparing M$_{JHK}$ for 2M0355 to the predicted values for an equivalent spectral type object based on the \citet{Faherty12} polynomial, we find it to be [-0.9,-0.5,-0.1] magnitude underluminous at M$_{J}$,M$_{H}$, and M$_{K}$, respectively.   As noted in \citet{Faherty12}, the population of low surface gravity L dwarfs is consistently red and underluminous--by up to 1.0 mag in M$_{JHK}$-- compared to equivalent spectral type objects. As demonstrated in Figure ~\ref{fig:spt}, 2M0355 clearly follows this trend.  As discussed in \citet{Faherty12} evolutionary models trace low-surface gravity objects at temperatures several hundred degrees lower than expected for equivalent spectral type objects on near-IR color magnitude diagrams, providing a potential explanation for the deviation in absolute magnitudes of low-gravity L dwarfs.  Extending this analysis to 2M0355 we conclude that one explanation for its peculiar near-IR absolute magnitudes is that this source is cooler than normal L5 field dwarfs.

\begin{figure*}[!ht]
\begin{center}
\epsscale{1.0}
\plotone{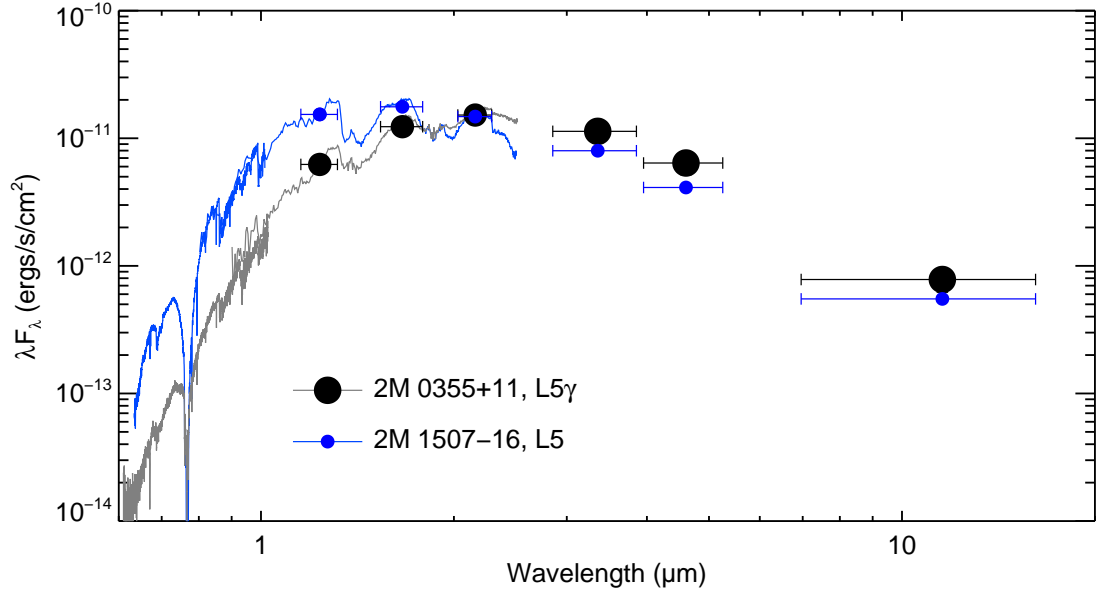}
\end{center}
\caption{The optical spectra, 2MASS $JHK$, and WISE $W1,W2,W3,W4$ photometry of 2M0355 (grey spectra and black filled circles) compared to the field L5 2M1507 (blue spectra and filled points).  The absolute photometry calculated from the parallax of each object (this work; \citealt{Dahn02}) as well as flux-calibrated optical spectra scaled to the near-IR photometry are transformed into $\lambda$F$_{\lambda}$. 2M0355 is underluminous through $K$ and then overluminous through $\sim$12 $\mu$m.} 
\label{fig:SED}
\end{figure*}

In Figure~\ref{fig:SED} we compare the full spectral energy distribution (SED) of 2M0355 to the field L5 dwarf 2MASSJ1507476-162738 (2M1507-\citealt{Reid00}; \citealt{Dahn02}). Combining the optical spectra, MKO $JHK$, and WISE $W1,W2,W3$ absolute photometry for each we confirm that the SED for 2M0355 is underluminous compared to the field object through $K$ band.  However, redward of $\sim$ 2.5$\mu$m, 2M0355 switches to being overluminous through at least 12 $\mu$m.   Following the method described in \citet{Cushing05}, we combine the flux-calibrated optical and near-IR spectra as well as WISE photometry and calculate bolometric luminosities for both 2M0355 and 2M1507.  We linearly interpolate between the centers of each WISE bandpass (W1: 3.4; W2: 4.6; W3: 11.6) and assume a Rayleigh-Jeans tail  for $\lambda>$11.6 $\mu$m.  We find that 2M0355 is slightly more luminous than 2M1507 by $\Delta$ log$_{10}$ (L$_{2M0355}$/L$_{2M1507}$)=0.12$\pm$0.1.   The overall luminosity of our source is further evidence that it is young, and we surmise that enhanced photospheric dust which weakens molecular bands and shifts flux to longer wavelengths is the most likely cause of the red SED.

\begin{figure*}[!ht]
\begin{center}
\epsscale{1.0}
\plotone{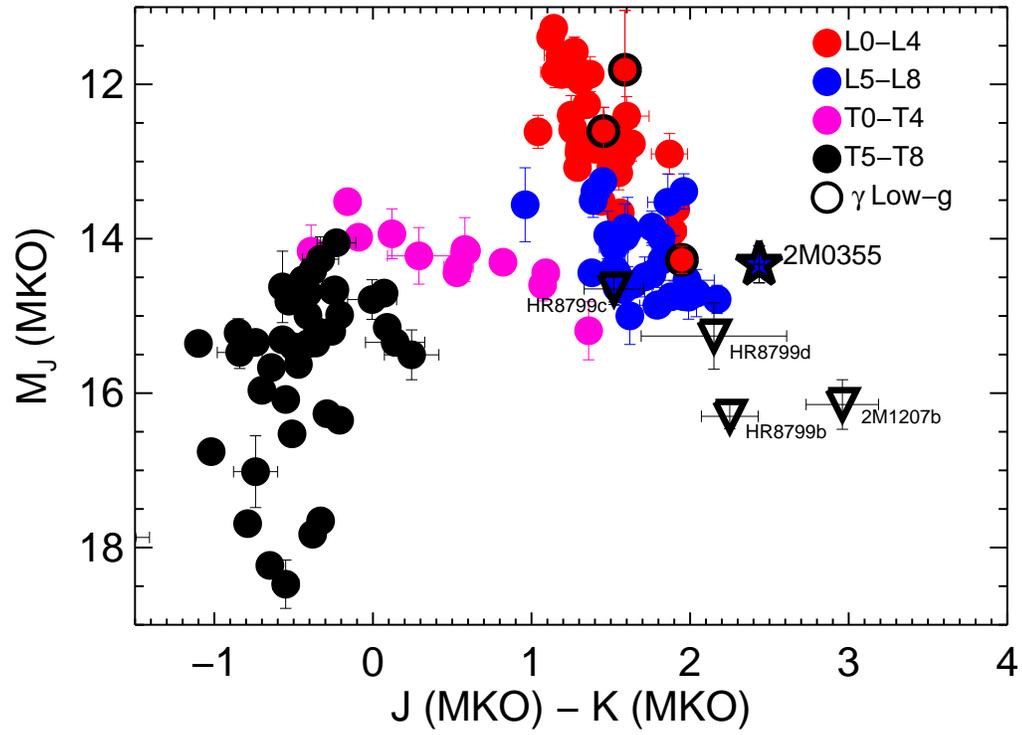}
\end{center}
\caption{The near-infrared color magnitude diagram for field and  low surface gravity L$\gamma$ brown dwarfs as well as giant planetary mass companions. Absolute magnitudes were derived from parallaxes reported in \citet{Faherty12} and \citet{Dupuy12}. We highlight the location of 2M0355 and note that it occupies a similar region of color-magnitude space as 2M1207b and HR8799b. } 
\label{fig:TM0355}
\end{figure*}

In Figure~\ref{fig:TM0355} we show the near-infrared color-magnitude diagram for the field brown dwarf population (color-coded by spectral type), 2M1207b, the HR8799bcd planets, and 2M0355.  The low luminosity and extremely red $J-K$ color of 2M0355 place it at the red edge of the brown dwarf population, in a similar region as 2M1207b. \citet{Barman11} find the positions of the giant exoplanets on this color-magnitude diagram--which are also redward and underluminous of the brown dwarf population--can be reproduced by allowing low T$_{eff}$ models (typically assumed cloud-free) to have clouds extending across their photospheres (see also \citealt{Bowler10}; \citealt{Currie11}; \citealt{Hinz10}; \citealt{Marley12}; \citealt{Madhusudhan11};  \citealt{Skemer12}). 2M1207b and HR8799bcd are young ($\sim$10 Myr and 30-160 Myr; respectively: \citealt{Chauvin04}; \citealt{Marois08}; \citealt{Marois10}) so youth is thought to be correlated with enhanced photospheric dust among the low-luminosity, low-temperature brown dwarfs and giant exoplanets (see also \citealt{Burgasser10}; \citealt{Faherty12}). 

Consequently, the position of 2M0355 on Figure ~\ref{fig:TM0355} leads us to conclude that in agreement with indications from the SED in Figure ~\ref{fig:SED} this source is both young and dusty.  

\section{Conclusions}
2M0355 is the reddest isolated L dwarfs yet characterized in the near- and mid-infrared. \citet{Cruz09} classified 2M0355 as L5$\gamma$, indicating low surface gravity spectral signatures. The similarity of the near-infrared spectrum to that of the $\sim$10 Myr planetary-mass object 2M1207b supports the conclusion that the object is young. Furthermore, a comparison with the near and mid-infrared colors of the known population of low surface gravity or L$\gamma$ dwarfs demonstrates that 2M0355 is the most extreme example of this class currently known. 

Combining optical spectra and absolute near to mid-IR photometry, we compared the full spectral energy distribution of 2M0355 with the field L5 dwarf 2M1507-16.  We find that 2M0355 is underluminous in optical through $K$ band then switches to overluminous through at least 12$\mu$m compared to 2M1507-16. Calculating the bolometric luminosity by integrating over the optical and near-IR spectra as well as WISE photometry, shows that the overall luminosity of 2M0355 is overluminous compared to the field object.  We conclude that enhanced photospheric dust,  thought to be correlated with young, low-temperature, low-luminosity brown dwarfs and giant exoplanets, shifts flux to longer wavelengths creating the red SED.  The position of 2M0355 on the near-IR color magnitude diagram supports this conclusion as it appears redward and underluminous of the known population in a similar region as 2M1207b and HR8799bcd.  

Combining new proper motion and parallax measurements we calculate UVW velocities to evaluate membership in nearby young moving groups. We find the kinematics consistent with the young thin disk and the UV velocities for 2M0355 place it in a busy part of velocity space for young objects. A careful kinematic comparison with nearby young groups and the field population leads us to conclude that 2M0355 has a 42\% chance of membership in AB Doradus.  2M0355 remains the brightest isolated  low surface gravity L dwarf studied to date and will prove to be a useful comparative object in low-temperature atmosphere studies directly applicable to giant exoplanets.

Despite the spectral similarity to 2M1207b in $H$ and $K$, 2M0355 is substantially different from the planetary-mass object in $J$band. This, combined with the older age estimate for 2M0355, cause the temperature and mass of 2M0355 to remain ambiguous. Nevertheless, we can use the object's absolute photometry and constrained age (assuming membership in AB~Doradus) to estimate these key properties. Using the evolutionary tracks for young, low mass objects of \citet{Baraffe02}, we estimate an effective temperature of $/sim$1500~K and a mass of $/sim$13 M$_{Jup}$ for an age of 50~Myr (the lower limit for the age of AB~Doradus). At the upper age limit for AB~Doradus, $/sim$150~Myr, the mass of 2M0355 would be closer to $/sim$30~M$_{Jup}$. As a field object, the absolute magnitudes of 2M0355 correspond to an object of $\sim$70~M$_{Jup}$, slightly below hydrogen burning minimum mass.

\acknowledgments{We acknowledge receipt of observation time through NOAO and we would like to thank 4.0m telescope operators C. Aguilera, M. Gonzalez, and A. Alvarez. We also thank M. Cushing and the anonymous referee for useful comments regarding the manuscript. KC gratefully acknowledges support from the Research Initiative for Scientific Enhancement program at Hunter College funded by the National Institute of Health. This publication has made use of the data products from the Two Micron All-Sky Survey, which is a joint project of the University of Massachusetts and the Infrared Processing and Analysis Center/California Institute of Technology, funded by the National Aeronautics and Space Administration and the National Science Foundation. This research has made use of the NASA/ IPAC Infrared Science Archive, which is operated by the Jet Propulsion Laboratory, California Institute of Technology, under contract with the National Aeronautics and Space Administration. This research has benefitted from the M, L, and T dwarf compendium housed at DwarfArchives.org and maintained by Chris Gelino, Davy Kirkpatrick, and Adam Burgasser.  This research has made use of NASA's Astrophysics Data System. The authors wish to recognize and acknowledge the very significant cultural role and reverence that the summit of Mauna Kea has always had within the indigenous Hawaiian community.  We are most fortunate to have the opportunity to conduct observations from this mountain. }

\bibliographystyle{apj}
\bibliography{paper2}

\clearpage

\clearpage

\clearpage

\clearpage

\clearpage

\clearpage

\clearpage

\clearpage

\clearpage

\clearpage

\clearpage

\clearpage

\begin{deluxetable*}{lcccrrrrrrrrrrr}
\label{tab:tab1}
\tabletypesize{\scriptsize}
\tablecaption{ Near-Infrared Observations\label{observing}}
\tablewidth{0pt}
\tablehead{
\colhead{Telescope} &
\colhead{Instrument} &
\colhead{Exp Time x coadds} &
\colhead{Images} &
\colhead{Date} &
\colhead{Airmass} &
\colhead{Filter/Mode} \\
 & &
\colhead{(s)} &
&
 &
 &\\
\colhead{(1)} &
\colhead{(2)} &
\colhead{(3)} &
\colhead{(4)} &
\colhead{(5)} &
\colhead{(6)}
}
\startdata
CTIO 4M & ISPI & 30x2 & 5 & 2008 October 11 & 1.3 & $J$ \\ 
 		&   & 10x4 & 5 & 2008 December 12 & 1.3 & $J$ \\ 
 		&   & 10x4 & 5 & 2009 November 30 & 1.3 &  $J$ \\ 
 		&   & 10x4 & 5 & 2010 January 28 & 1.5 &  $J$ \\ 
 		&   & 30x4 & 10 & 2011 November 11 & 1.3 & $J$ \\ 
 		&   & 30x2 & 10 & 2012 January 03 & 1.3 &  $J$ \\ 
		&   & 30x4 & 5   & 2012 February 05 & 1.4 & $J$ \\
		\hline
IRTF 	& SpeX& 90x1	 &10 &2011 December 7				 & 1.2  & Prism\\ 
 	 	& SpeX&300x1 &6 &2007 November 13				 &1.0   & SXD\\ 
\enddata
\end{deluxetable*}
\clearpage

\begin{deluxetable*}{ccccccccccccccr}
\label{tab:tab1}
\tabletypesize{\scriptsize}
\tablecaption{Average Near-Infrared and Mid-Infrared Colors of L dwarfs\label{meancolors}}
\tablewidth{0pt}
\tablehead{
\colhead{SpT} &
\colhead{N$_{WISE}$\tablenotemark{a}} &
\colhead{N$_{2MASS}$\tablenotemark{a}} &
\colhead{N$_{Low-G}$} &
\colhead{(J-K$_{s}$)$_{avg}$} &
\colhead{$\sigma$(J-K$_{s}$)} &
\colhead{(W1-W2)$_{avg}$} &
\colhead{$\sigma$(W1-W2)} \\
\colhead{(1)} &
\colhead{(2)} &
\colhead{(3)} &
\colhead{(4)} &
\colhead{(5)} &
\colhead{(6)} &
\colhead{(7)} &
\colhead{(8)} \\
}
\startdata
   L0    &  143    &  102    &   11    &      1.30    &      0.15    &      0.27    &      0.06   \\
   L1    &  125    &   95    &    2    &      1.35    &      0.16    &      0.26    &      0.06   \\
   L2    &   58    &   60    &    3    &      1.48    &      0.17    &      0.28    &      0.07   \\
   L3    &   69    &   51    &    3    &      1.64    &      0.18    &      0.31    &      0.06   \\
   L4    &   37    &   33    &    5    &      1.69    &      0.24    &      0.34    &      0.08   \\
   L5    &   43    &   28    &    2    &      1.72    &      0.22    &      0.35    &      0.08   \\
   L6    &   25    &   13    &    0    &      1.84    &      0.25    &      0.42    &      0.11   \\
   L7    &   13    &    9    &    0    &      1.75    &      0.26    &      0.46    &      0.09   \\
 L8-9    &   19    &   10    &    0    &      1.85    &      0.17    &      0.54    &      0.08   \\
\hline
\enddata
\tablenotetext{a}{Only normal (non- low surface gravity, subdwarf, or young) L dwarfs with photometric uncertainty $<$ 0.1 were used in calculating the average.}
\end{deluxetable*}

\clearpage
\begin{landscape}
\begin{deluxetable*}{lcccccccccccccr}
\label{tab:tab2}
\tabletypesize{\scriptsize}

\tablecaption{Photometric Properties of  low surface Gravity L$\gamma$ Dwarfs\label{low-G}}
\tablewidth{0pt}
\tablehead{
\colhead{Name} &
\colhead{SpT (OpT)} &
\colhead{J\tablenotemark{a}} &
\colhead{H\tablenotemark{a}} &
\colhead{K$_{s}$\tablenotemark{a}} &
\colhead{W1\tablenotemark{a}} &
\colhead{W2\tablenotemark{a}} &
\colhead{W3\tablenotemark{a}} &
\colhead{W4\tablenotemark{a}} &
\colhead{Ref}\\
\colhead{(1)} &
\colhead{(2)} &
\colhead{(3)} &
\colhead{(4)} &
\colhead{(5)} &
\colhead{(6)} &
\colhead{(7)} &
\colhead{(8)} &
\colhead{(9)} &
\colhead{(10)} \\
}
\startdata

2MASSJ003255.84-440505.8       &   L0.0$\gamma$    &            14.78   $\pm$      0.03    &     13.86   $\pm$      0.03    &     13.27   $\pm$      0.04    &     12.82   $\pm$      0.03    &     12.49   $\pm$      0.03    &     11.73   $\pm$      0.19    &      9.29   $\pm$   null       &    1,2\\
2MASSJ003743.06-584622.9       &   L0.0$\gamma$    &            15.37   $\pm$      0.05    &     14.26   $\pm$      0.05    &     13.59   $\pm$      0.04    &     13.13   $\pm$      0.03    &     12.74   $\pm$      0.03    &     12.56   $\pm$      0.38    &      9.32   $\pm$   null       &    1,2\\
2MASSJ012445.99-574537.9       &   L0.0$\gamma$    &            16.31   $\pm$      0.10    &     15.06   $\pm$      0.09    &     14.32   $\pm$      0.09    &     13.77   $\pm$      0.03    &     13.34   $\pm$      0.03    &     12.45   $\pm$      0.31    &      8.91   $\pm$   null       &    1,2\\
2MASSJ014158.23-463357.4       &   L0.0$\gamma$    &            14.83   $\pm$      0.04    &     13.88   $\pm$      0.02    &     13.10   $\pm$      0.03    &     12.55   $\pm$      0.02    &     12.17   $\pm$      0.02    &     11.92   $\pm$      0.21    &      9.24   $\pm$   null       &    3,2\\
2MASSJ022354.64-581506.7       &   L0.0$\gamma$    &            15.07   $\pm$      0.05    &     14.00   $\pm$      0.04    &     13.42   $\pm$      0.04    &     12.82   $\pm$      0.02    &     12.43   $\pm$      0.02    &     11.64   $\pm$      0.15    &      9.47   $\pm$   null       &    1,2\\
2MASSJ023400.93-644206.8       &   L0.0$\gamma$    &            15.33   $\pm$      0.06    &     14.44   $\pm$      0.06    &     13.85   $\pm$      0.07    &     13.25   $\pm$      0.03    &     12.91   $\pm$      0.03    &     12.62   $\pm$      0.28    &      9.49   $\pm$   null       &    4\\
2MASSJ024111.51-032658.7       &   L0.0$\gamma$    &            15.80   $\pm$      0.06    &     14.81   $\pm$      0.05    &     14.04   $\pm$      0.05    &     13.64   $\pm$      0.03    &     13.26   $\pm$      0.03    &     12.77   $\pm$      0.42    &      9.00   $\pm$   null       &    2,5\\
2MASSJ032310.02-463123.7       &   L0.0$\gamma$    &            15.39   $\pm$      0.07    &     14.32   $\pm$      0.06    &     13.70   $\pm$      0.05    &     13.08   $\pm$      0.02    &     12.67   $\pm$      0.02    &     11.94   $\pm$      0.16    &      9.18   $\pm$   null       &    1,2\\
2MASSJ040626.77-381210.2       &   L0.0$\gamma$    &            16.77   $\pm$      0.13    &     15.71   $\pm$      0.10    &     15.11   $\pm$      0.12    &     14.45   $\pm$      0.03    &     14.10   $\pm$      0.04    &     12.52   $\pm$   null       &      9.10   $\pm$   null       &    4\\
2MASSJ195647.00-754227.0       &   L0.0$\gamma$    &            16.15   $\pm$      0.10    &     15.04   $\pm$      0.10    &     14.23   $\pm$      0.07    &     13.69   $\pm$      0.03    &     13.25   $\pm$      0.03    &     12.68   $\pm$   null       &      9.17   $\pm$   null       &    1,2\\
2MASSJ221344.91-213607.9       &   L0.0$\gamma$    &            15.38   $\pm$      0.03    &     14.40   $\pm$      0.06    &     13.76   $\pm$      0.04    &     13.23   $\pm$      0.03    &     12.83   $\pm$      0.03    &     11.55   $\pm$      0.20    &      9.07   $\pm$   null       &    2,5\\
2MASSJ000402.88-641035.8       &   L1.0$\gamma$    &            15.79   $\pm$      0.07    &     14.83   $\pm$      0.07    &     14.01   $\pm$      0.05    &     13.37   $\pm$      0.03    &     12.94   $\pm$      0.03    &     12.18   $\pm$      0.24    &      9.16   $\pm$   null       &    4\\
2MASSJ051846.16-275645.7       &   L1.0$\gamma$    &            15.26   $\pm$      0.04    &     14.30   $\pm$      0.05    &     13.62   $\pm$      0.04    &     13.05   $\pm$      0.02    &     12.66   $\pm$      0.03    &     12.58   $\pm$      0.35    &      9.22   $\pm$   null       &    5,6\\
2MASSJ030320.42-731230.0       &   L2.0$\gamma$    &            16.14   $\pm$      0.11    &     15.10   $\pm$      0.09    &     14.32   $\pm$      0.08    &     13.78   $\pm$      0.03    &     13.35   $\pm$      0.03    &     12.29   $\pm$      0.17    &      9.34   $\pm$      0.34    &    4\\
2MASSJ053619.98-192039.6       &   L2.0$\gamma$    &            15.77   $\pm$      0.07    &     14.69   $\pm$      0.07    &     13.85   $\pm$      0.06    &     13.26   $\pm$      0.03    &     12.79   $\pm$      0.03    &     12.55   $\pm$      0.40    &      9.24   $\pm$   null       &    5,6\\
2MASSJ232252.99-615127.5       &   L2.0$\gamma$    &            15.55   $\pm$      0.06    &     14.54   $\pm$      0.06    &     13.86   $\pm$      0.04    &     13.24   $\pm$      0.03    &     12.84   $\pm$      0.03    &     12.68   $\pm$      0.39    &      9.38   $\pm$   null       &    1,2\\
2MASSJ172600.07+153819.0      &   L3.5$\gamma$    &       15.67   $\pm$      0.06    &     14.47   $\pm$      0.05    &     13.66   $\pm$      0.05    &     13.07   $\pm$      0.03    &     12.69   $\pm$      0.03    &     11.56   $\pm$      0.16    &      9.31   $\pm$   null       &    2,7\\
2MASSJ212650.40-814029.3       &   L3.0$\gamma$    &            15.54   $\pm$      0.06    &     14.41   $\pm$      0.05    &     13.55   $\pm$      0.04    &     12.91   $\pm$      0.02    &     12.47   $\pm$      0.02    &     11.89   $\pm$      0.16    &      9.36   $\pm$   null       &    1,2\\
2MASSJ220813.63+292121.5      &   L3.0$\gamma$    &            15.80   $\pm$      0.08    &     14.79   $\pm$      0.07    &     14.15   $\pm$      0.07    &     13.35   $\pm$      0.03    &     12.89   $\pm$      0.03    &     12.58   $\pm$      0.39    &      9.30   $\pm$   null       &    2,7\\
2MASSJ012621.09+142805.7      &   L4.0$\gamma$    &            17.11   $\pm$      0.21    &     16.17   $\pm$      0.22    &     15.28   $\pm$      0.15    &     14.24   $\pm$      0.03    &     13.70   $\pm$      0.04    &     12.38   $\pm$   null       &      9.13   $\pm$   null       &     6,8\\
2MASSJ050124.06-001045.2       &   L4.0$\gamma$    &            14.98   $\pm$      0.04    &     13.71   $\pm$      0.03    &     12.96   $\pm$      0.03    &     12.05   $\pm$      0.02    &     11.52   $\pm$      0.02    &     10.95   $\pm$      0.11    &      9.17   $\pm$   null       &    1,2\\
2MASSJ155152.37+094114.8      &   L4.0$\gamma$    &            16.32   $\pm$      0.11    &     15.11   $\pm$      0.07    &     14.31   $\pm$      0.06    &     13.60   $\pm$      0.03    &     13.12   $\pm$      0.03    &     12.68   $\pm$      0.48    &      9.16   $\pm$   null       &    1,6\\
2MASSJ161542.55+495321.1      &   L4.0$\gamma$    &            16.79   $\pm$      0.14    &     15.33   $\pm$      0.10    &     14.31   $\pm$      0.07    &     13.20   $\pm$      0.02    &     12.62   $\pm$      0.02    &     12.13   $\pm$      0.13    &      9.31   $\pm$   null       &    5,6\\
2MASSJ224953.45+004404.6      &   L4.0$\gamma$    &            16.59   $\pm$      0.12    &     15.42   $\pm$      0.11    &     14.36   $\pm$      0.07    &     13.58   $\pm$      0.03    &     13.14   $\pm$      0.05    &     11.28   $\pm$   null       &      7.69   $\pm$   null       &    6,9,10,11\\
2MASSJ035523.37+113343.7     &   L5.0$\gamma$    &             14.05   $\pm$      0.02    &     12.53   $\pm$      0.03    &     11.53   $\pm$      0.02    &     10.53   $\pm$      0.02    &      9.94   $\pm$      0.02    &      9.29   $\pm$      0.04    &      8.32   $\pm$   null       &    1	,2\\
2MASSJ042107.18-630602.2      &   L5.0$\gamma$    &             15.57   $\pm$      0.05    &     14.28   $\pm$      0.04    &     13.45   $\pm$      0.04    &     12.56   $\pm$      0.02    &     12.14   $\pm$      0.02    &     11.60   $\pm$      0.10    &      9.25   $\pm$   null       &    2,5\\

\hline
\enddata
\tablenotetext{a}{$JHK_{s}$ photometry from the Two Micron All Sky Catalog (\citealt{Skrutskie06}) and the $W1,W2,W3,W4$ from the Wide-field Infrared Survey Explorer (\citealt{Wright10})}
\tablecomments{References: (1) \citet{Reid08} (2) \citet{Cruz09} (3) \citet{Kirkpatrick06} (4) \citet{Kirkpatrick10} (5) \citet{Cruz07} (6) Cruz et al. in prep (7) \citet{Kirkpatrick00} (8) \citet{Metchev08} (9) \citet{Geballe02} (10) \citet{Hawley02} (11) \citet{Nakajima04}}

\end{deluxetable*}
\clearpage
\end{landscape}

\begin{deluxetable*}{ccccccccccccccr}
\label{tab:tab2}
\tabletypesize{\scriptsize}
\tablecaption{Colors of  low surface Gravity L$\gamma$ Dwarfs\label{colorslow-G}}
\tablewidth{0pt}
\tablehead{
\colhead{Name} &
\colhead{SpT} &
\colhead{(J-K$_{s}$)} &
\colhead{(W1-W2)} &
\colhead{$\Delta_{(J-K_{s})}$\tablenotemark{a}} &
\colhead{$\Delta_{(W1-W2)}$\tablenotemark{a}} \\
\colhead{2MASS} &
\colhead{OpT} &
\colhead{2MASS} &
\colhead{WISE} \\
\colhead{(1)} &
\colhead{(2)} &
\colhead{(3)} &
\colhead{(4)} &
\colhead{(5)} &
\colhead{(6)} \\
}
\startdata
     2MASSJ0032-4405    &   L0.0$\gamma$    &      1.51   $\pm$      0.05    &      0.33   $\pm$      0.04    &      0.21    &      0.06   \\
     2MASSJ0037-5846    &   L0.0$\gamma$    &      1.78   $\pm$      0.06    &      0.39   $\pm$      0.04    &      0.48    &      0.12   \\
     2MASSJ0124-5745    &   L0.0$\gamma$    &      1.99   $\pm$      0.13    &      0.43   $\pm$      0.04    &      0.69    &      0.16   \\
     2MASSJ0141-4633    &   L0.0$\gamma$    &      1.73   $\pm$      0.05    &      0.38   $\pm$      0.03    &      0.43    &      0.11   \\
     2MASSJ0223-5815    &   L0.0$\gamma$    &      1.65   $\pm$      0.06    &      0.39   $\pm$      0.03    &      0.35    &      0.12   \\
     2MASSJ0234-6442    &   L0.0$\gamma$    &      1.48   $\pm$      0.09    &      0.34   $\pm$      0.04    &      0.18    &      0.07   \\
     2MASSJ0241-0326    &   L0.0$\gamma$    &      1.76   $\pm$      0.08    &      0.38   $\pm$      0.04    &      0.46    &      0.11   \\
     2MASSJ0323-4631    &   L0.0$\gamma$    &      1.69   $\pm$      0.09    &      0.41   $\pm$      0.03    &      0.39    &      0.14   \\
     2MASSJ0406-3812    &   L0.0$\gamma$    &      1.66   $\pm$      0.18    &      0.35   $\pm$      0.05    &      0.36    &      0.08   \\
     2MASSJ1956-7542    &   L0.0$\gamma$    &      1.92   $\pm$      0.12    &      0.44   $\pm$      0.04    &      0.62    &      0.17   \\
     2MASSJ2213-2136    &   L0.0$\gamma$    &      1.62   $\pm$      0.05    &      0.40   $\pm$      0.04    &      0.32    &      0.13   \\
     2MASSJ0004-6410    &   L1.0$\gamma$    &      1.78   $\pm$      0.09    &      0.43   $\pm$      0.04    &      0.43    &      0.17   \\
     2MASSJ0518-2756    &   L1.0$\gamma$    &      1.64   $\pm$      0.06    &      0.39   $\pm$      0.04    &      0.29    &      0.13   \\
     2MASSJ0303-7312    &   L2.0$\gamma$    &      1.82   $\pm$      0.14    &      0.43   $\pm$      0.04    &      0.34    &      0.15   \\
     2MASSJ0536-1920    &   L2.0$\gamma$    &      1.92   $\pm$      0.09    &      0.47   $\pm$      0.04    &      0.44    &      0.19   \\
     2MASSJ2322-6151    &   L2.0$\gamma$    &      1.69   $\pm$      0.07    &      0.40   $\pm$      0.04    &      0.21    &      0.12   \\
     2MASSJ1726+1538    &   L3.5$\gamma$    &      2.01   $\pm$      0.08    &      0.38   $\pm$      0.04    &      0.37    &      0.07   \\
     2MASSJ2126-8140    &   L3.0$\gamma$    &      1.99   $\pm$      0.07    &      0.44   $\pm$      0.03    &      0.35    &      0.13   \\
     2MASSJ2208+2921    &   L3.0$\gamma$    &      1.65   $\pm$      0.11    &      0.47   $\pm$      0.04    &      0.01    &      0.16   \\
     2MASSJ0126+1428    &   L4.0$\gamma$    &      1.83   $\pm$      0.26    &      0.54   $\pm$      0.05    &      0.14    &      0.20   \\
     2MASSJ0501-0010    &   L4.0$\gamma$    &      2.02   $\pm$      0.05    &      0.53   $\pm$      0.03    &      0.33    &      0.19   \\
     2MASSJ1551+0941    &   L4.0$\gamma$    &      2.01   $\pm$      0.12    &      0.48   $\pm$      0.04    &      0.32    &      0.14   \\
     2MASSJ1615+4953    &   L4.0$\gamma$    &      2.48   $\pm$      0.16    &      0.58   $\pm$      0.03    &      0.79    &      0.24   \\
     2MASSJ2249+0044    &   L4.0$\gamma$    &      2.23   $\pm$      0.14    &      0.43   $\pm$      0.06    &      0.54    &      0.09   \\     
     2MASSJ0355+1133    &   L5.0$\gamma$    &      2.52   $\pm$      0.03    &      0.59   $\pm$      0.03    &      0.80    &      0.24   \\
     2MASSJ0421-6306    &   L5.0$\gamma$    &      2.12   $\pm$      0.06    &      0.42   $\pm$      0.03    &      0.40    &      0.07   \\
\hline
\enddata
\tablenotetext{a}{$\Delta$ values are calculated from the mean colors listed  in Table ~\ref{meancolors}.}
\end{deluxetable*}
\clearpage

\clearpage

\begin{deluxetable*}{lll}
\label{tab:tab5}
\tabletypesize{\scriptsize}
\tablecaption{Properties of 2MASSJ035523.37+113343.7\label{properties}}
\tablewidth{0pt}
\tablehead{
\colhead{Parameter} &
\colhead{Value} &
\colhead{Reference} \\
\colhead{(1)} &
\colhead{(2)} &
\colhead{(3)} \\
}
\startdata
Parameter & Value & Reference\\
\hline
RA,Dec (J2000) & 03$^{h}$55$^{m}$23.37$^{s}$ +11$^{\circ}$33$^{`}$43.7$^{"}$ & 1\\
Optical SpT & L5$\gamma$&2\\
J (2MASS)&14.05$\pm$0.02&1\\
H (2MASS)&12.53$\pm$0.03&1\\
K$_{s}$ (2MASS)&11.53$\pm$0.02&1\\
J (MKO)\tablenotemark{a}&13.90$\pm$0.03 &4\\
H (MKO)\tablenotemark{a}&12.60$\pm$0.03&4\\
K (MKO)\tablenotemark{a}&11.46$\pm$0.02&4\\
M$_{J}$ (MKO)&14.33$\pm$0.24 &4 \\
M$_{H}$ (MKO)&13.03$\pm$0.24 &4 \\
M$_{K}$ (MKO)&11.89$\pm$0.23 &4\\
W1&10.53$\pm$0.02&3\\
W2&9.94$\pm$  0.02&3\\
W3&9.29$\pm$  0.04&3\\
W4&8.32$\pm$  null&3\\
$\mu_{\alpha}$&218$\pm$  5 mas yr$^{-1}$&4\\
$\mu_{\delta}$& -626$\pm$  5 mas yr$^{-1}$&4\\
$\pi_{abs}$&122$\pm$  13 mas &4\\
RV&11.92$\pm$0.22 km s$^{-1}$&5\\
U\tablenotemark{b}&-5.9$\pm$1.5 km s$^{-1}$&4\\
V\tablenotemark{b}&-23.6$\pm$2.0 km s$^{-1}$&4\\
W\tablenotemark{b}&-14.6$\pm$1.3 km s$^{-1}$&4\\
X\tablenotemark{b}&-7.0$\pm$0.7~pc&4\\
Y\tablenotemark{b}&0.2$\pm$0.4~pc&4\\
Z\tablenotemark{b}&-4.2$\pm$0.4~pc&4\\
Age\tablenotemark{c} & 50-150 Myr & 4\\
Mass\tablenotemark{c} &13 - 30 M$_{Jup}$ & 4 \\
\hline
\enddata
\tablenotetext{a}{MKO values calculated using the transformations in \cite{Stephens04}}
\tablenotetext{b}{UVW and XYZ values are calculated in a left-handed coordinate system with U and X positive toward the Galactic center.}
\tablenotetext{c} {Age and mass are based on the fact that we estimate a 43\% probability that 2M0355 is an AB Doradus member.}

\tablecomments{References: (1) \citet{Cutri03} (2) \citet{Cruz09} (3) \citet{Wright10} (4) This Paper (5) \citet{Blake10} }
\end{deluxetable*}
\clearpage

\end{document}